\documentclass{aa} 

\usepackage{txfonts}
\usepackage{natbib}
\usepackage{graphicx}
\usepackage{color}
\usepackage{soul}

\definecolor{red}{rgb}{1,0,0}
\definecolor{blue}{rgb}{0,0,1}
\definecolor{green}{rgb}{0,1,0}

\newcommand{\corr}{}
\newcommand{\correq}{}

\bibpunct{(}{)}{;}{a}{}{,}

\title{Post-coronagraphic tip-tilt sensing for vortex phase masks:\\the QACITS technique}
\titlerunning{The QACITS technique}

\author{
   E.~Huby\inst{1}
\and P.~Baudoz \inst{2}
\and D.~Mawet \inst{3,4}
\and O.~Absil \inst{1}\fnmsep\thanks{F.R.S.-FNRS Research Associate}
}
\institute{
	D\'epartement d’Astrophysique, G\'eophysique et Oc\'eanographie, Universit\'e de Liège, 19 All\'ee du Six Ao\^ut, 4000 Li\`ege, Belgium
	\and LESIA, Observatoire de Paris, CNRS, UPMC, Universit\'e Paris-Diderot, Paris Sciences et Lettres, 5 place Jules Janssen, 92195 Meudon, France
	\and Department of Astronomy, California Institute of Technology, 1200 E. California Blvd., Pasadena, CA 91125, USA
	\and Jet Propulsion Laboratory, 4800 Oak Grove Drive, Pasadena CA 91109, USA
}

\date{Received ... / Accepted ...}

\abstract
{
Small inner working angle coronagraphs, like the vortex phase mask, are essential to exploit the full potential of ground-based telescopes in the context of exoplanet detection and characterization. However, the drawback of this attractive feature is a high sensitivity to pointing errors, which degrades the performance of the coronagraph.
}
{
We propose a tip-tilt retrieval technique based on the analysis of the final coronagraphic image, hereafter called Quadrant Analysis of Coronagraphic Images for Tip-tilt Sensing (QACITS). 
}
{
Under the assumption of small phase aberrations, we show that the behaviour of the vortex phase mask can be simply described from the entrance pupil to the Lyot stop plane by Zernike polynomials. This convenient formalism is used to establish the theoretical basis of the QACITS technique. Simulations have been performed to demonstrate the validity and limits of the technique, including the case of a centrally obstructed pupil.
}
{
The QACITS technique principle is validated by experimental results in the case of an unobstructed circular aperture, and by simulations in presence of a central obstruction. The typical configuration of the Keck telescope (24\% central obstruction) has been simulated with additional high order aberrations. In these conditions, our simulations show that the QACITS technique is still adapted to centrally obstructed pupils and performs tip-tilt retrieval with a precision of $5 \times 10^{-2}\lambda/D$ when wavefront errors amount to $\lambda/14$\,rms and $10^{-2}\lambda/D$ for $\lambda/70$\,rms errors (with $\lambda$ the wavelength and $D$ the pupil diameter).
}
{
We have developed and demonstrated a tip-tilt sensing technique for vortex coronagraphs. The implementation of the QACITS technique is based on the analysis of the scientific image and does not require any modification of the original setup. Current facilities equipped with a vortex phase mask can thus directly benefit from this technique to improve the contrast performance close to the axis.
}


\begin{document}

\maketitle


\section{Introduction}

Vortex coronagraphs \citep[VC, ][]{Mawet2005, Foo2005, Mawet2009} stand amongst the most promising focal plane phase masks envisioned for the next generation instruments of future very large telescopes \citep[e.g. METIS, ][]{Brandl2014}. Theoretically, this coronagraphic solution provides a perfect star light rejection and other valuable features for direct imaging and characterization of exoplanets: achromaticity, continuous 360$^\circ$ discovery space and small inner working angle (angular distance where the off-axis transmission reaches 50\%). For these reasons, vortex phase masks already equip several infrared instruments on 10\,m class ground-based telescopes, namely VLT/NACO \citep{Mawet2013}, VLT/VISIR \citep{Delacroix2012, Kerber2014}, LBT/LMIRCam \citep{Defrere2014}, Subaru/SCExAO \corr{\citep{Jovanovic2015}} and very recently Keck/NIRC2. Scientific results have been obtained using the coronagraphic mode of these facilities, leading to the detection of exoplanets and circumstellar disks \citep[e.g.][]{Absil2013,Milli2014,Reggiani2014}. The off-axis well-corrected subaperture on the Palomar Hale telescope \citep{Serabyn2007} also provides a VC mode, which has led to impressive results \citep{Serabyn2010}, including the detection of a companion very close to its host star ($\epsilon$ Cephei), at an angular separation of $1.1\,\lambda/D$ \citep[][$\lambda$ and $D$ being the wavelength of observation and the telescope diameter respectively]{Mawet2011}.

However, a small inner working angle comes inevitably at a cost. Vortex phase masks, and in particular vortices of topological charge \corr{$l_p=2$} like the Annular Groove Phase Masks \citep[AGPM, ][]{Mawet2005}, are amongst the focal plane masks that offer the narrowest inner working angle (down to $1\,\lambda/D$), but it also means that they are highly sensitive to the centering of the star on the mask. Accurate tracking systems are therefore required to limit the contrast loss due to pointing errors. A variety of low-order aberration sensing techniques exists and is used in current instruments, as reviewed by \cite{Mawet2012}. In order to avoid non-common path errors, the sensor must be placed as close as possible to the coronagraphic phase mask. Solutions include sensors built just before the coronagraphic mask, like the Differential Tip-Tilt Sensor (DTTS) of SPHERE \citep[][where part of the light is diverted thanks to a dichroic plate]{Baudoz2010}, or the Cal low-order wavefront sensor of GPI \citep[][which uses the light rejected by the central spot of the occulting mask]{Wallace2010}. For focal-plane phase masks, the later solution cannot be implemented, but a comparable solution has been proposed by \cite{Singh2014}, making use of the light rejected by the coronagraph thanks to a reflective Lyot stop. Finally, phase retrieval techniques can also be applied directly from the image acquired by the scientific detector, like the COFFEE sensor implemented in the SPHERE instrument \citep{Sauvage2012}.

In this paper, we propose a solution belonging to the latter category. It is based on the analysis of the final image produced by a VC to retrieve the tip-tilt affecting the beam incident on the phase mask. The principle of this technique, referred to as Quadrant Analysis of Coronagraphic Images for Tip-tilt Sensing (QACITS) has first been introduced by \cite{Mas2012} for the four quadrant phase mask \citep[FQPM, ][]{Rouan2000}. It consists of quantifying the asymmetry observed in the coronagraphic point spread function (PSF), using the same principle as a quadrant cell detector. The differential intensities, or intensity gradients, are related to the pointing error and allow the estimation of the tip-tilt aberration affecting the beam. The simplicity of this technique makes it very easy to implement on current instruments working with a vortex phase mask, as there is no need for any modification of the optical setup.

In the next section, we describe the QACITS technique applied to the perfect VC with an unobstructed pupil, and in particular the mathematical model linking the asymmetry in the image and the tip-tilt, followed in Sect.\,\ref{sec:experimental_validation} by an experimental validation of the model. For the sake of clarity, the details of the analytical computation are given in the appendices, where we introduce a formalism based on Zernike polynomials. In Sect.\,\ref{sec:central_obstruction}, we detail the implications of a central obstruction on the PSF shape, and thus on the model used in QACITS. Additionally, we propose a slightly modified QACITS using two distinct image areas independently. In Sect.\,\ref{sec:aberrations}, we report on simulation results of the QACITS performance in presence of higher order aberrations affecting the wavefront. In the final section, we draw the conclusions of our study.

\section{QACITS: Quadrant Analysis of Coronagraphic Images for Tip-tilt Sensing}
\label{sec:qacits}

\begin{figure}
\centering
\includegraphics[width=.9\linewidth]{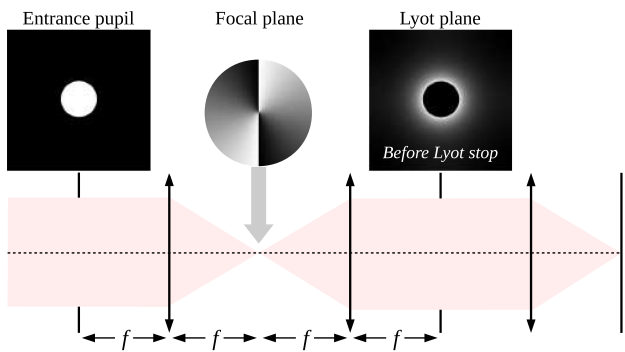}
\caption{Standard coronagraph layout, with a vortex phase mask at the focal plane and a Lyot stop at the second pupil plane (Lyot plane).}
\label{fig:layout}
\end{figure}

In this section, we introduce the QACITS post-coronagraphic technique to retrieve the tip-tilt affecting the beam upstream a vortex phase mask. The demonstration by \cite{Mas2012} in the case of the FQPM is based on simulations and experimental data but an analytical model could also be derived (\corr{P.\,Baudoz, }private communication). In the present section, we derive the analytical model for the VC of charge $l_p=2$, based on the typical coronagraph layout illustrated in Fig. \ref{fig:layout}. For that purpose, we use the Zernike formalism described in detail in Appendix \ref{app:zernike_analysis}.

\subsection{The quadrant analysis principle}

\begin{figure}
\centering
\includegraphics[width=.7\linewidth]{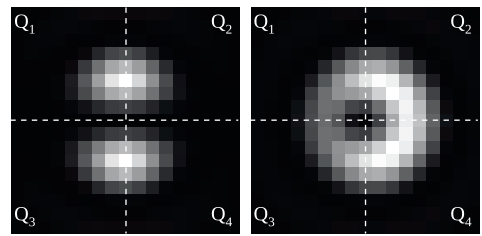}
\caption{Simulated images obtained for a tip-tilt of $0.2\lambda/D$ applied in the horizontal direction, for the case of a four-quadrant phase mask (left) and a vortex phase mask (right). Each quadrant $Q_i$ is a square of width  2 $\lambda/D$ and defines an area where the flux is integrated in order to quantify the asymmetry in the image.}
\label{fig:tilt_img}
\end{figure}

\cite{Mas2012} have shown that the amount of tip-tilt aberration that affects the wavefront upstream the coronagraphic mask can be retrieved by analysing the residuals of the attenuated on-axis image acquired by the scientific detector. Indeed, this aberration induces an asymmetry in the pattern, as illustrated in Fig.\,\ref{fig:tilt_img}. The asymmetry is quantified by two flux measurements, $\Delta I_x$ and $\Delta I_y$, corresponding to the flux gradient along two orthogonal directions in the image\corr{, which can be defined as}

\begin{equation}
\Delta I_x = \dfrac{(I_2+I_4) - (I_1+I_3)}{I_0} \text{ and } \Delta I_y = \dfrac{(I_1+I_2) - (I_3+I_4)}{I_0},
\end{equation}

\noindent with $I_k =\int_{Q_k}I$ the flux contained in each quadrant area $Q_k$ and $I_0 = \sum \limits_ i ^{4} \int_{Q_i} I_{\rm nc}$ the total amount of flux contained in the non coronagraphic image $I_{\rm nc}$. In practice, these areas are squares of width a few $\lambda/D$ ($2\lambda/D$ in Fig.~\ref{fig:tilt_img}). Empirically, \cite{Mas2012} found that in the small aberration approximation, these quantities are directly linked to the amount of tip-tilt in the $x$ and $y$ directions, $T_x$ and $T_y$ respectively, following the model

\begin{equation} 
\Delta I_x = \beta \left( T_x^3 + \alpha T_x T_y^2 \right) \text{ and } \Delta I_y = \beta \left( T_y^3 + \alpha T_y T_x^2 \right),
\label{eq:model_mas}
\end{equation}

\noindent $\beta$ being a normalization coefficient. In the case of the FQPM, they found $\alpha_{\mathrm{FQPM}}=4$ and managed to find an approximate solution of the system. In the following section, we show that for the vortex phase mask, the model can be derived analytically using the Zernike-based analysis.

\subsection{Zernike formalism: from entrance pupil to Lyot plane}

The phase of a tilted wavefront using Zernike polynomials expresses as

\begin{equation}
\phi = T_x Z_2(\vec{r}) + T_y Z_3(\vec{r}),
\label{eq:phase}
\end{equation}

\noindent with $\vec{r}=(r, \theta)$ the polar coordinates in the pupil plane, $T_x$ and $T_y$ the root mean square (rms) values for tip and tilt modes in radians, $Z_2 = 2 r \cos \theta$ and $Z_3 = 2 r \sin \theta$ the tip-tilt modes expressed as the standard Zernike polynomials described by \cite{Noll1976} and recalled in the Appendix \ref{app:zernike-polynomials}. These polynomials are normalized to 1\,rad\,rms. In the small aberration approximation, \corr{the exponential function describing the wavefront can be expanded and approximated by}

\begin{equation}
E_{\rm pup} = {\rm e}^{i \phi} \approx 1 + i \phi - \dfrac{\phi^2}{2}.
\end{equation}

\noindent The second order expansion is needed in order to make the non-symmetrical terms appear in the final PSF expression. The development of $\phi^2$ is a combination of the following terms, projected on the Zernike basis:

\begin{equation}
\begin{array}{rcl}
Z_2(\vec{r})^2& =& Z_1(\vec{r})+Z_4(\vec{r})/\sqrt{3} + 2 Z_6(\vec{r})/\sqrt{6}, \\
Z_3(\vec{r})^2 &= &  Z_1(\vec{r})+Z_4(\vec{r})/\sqrt{3} - 2 Z_6(\vec{r})/\sqrt{6},  \\
Z_2(\vec{r})Z_3(\vec{r}) &=& 2 Z_5(\vec{r}) / \sqrt{6}, \\
\end{array}
\end{equation}

\noindent \corr{where} $Z_1(\vec{r})$, $Z_4(\vec{r})$, $Z_5(\vec{r})$ and $Z_6(\vec{r})$ correspond respectively to the piston, focus and the two astigmatism modes (see Appendix \ref{app:zernike-polynomials} for the explanation about the numbering of the polynomials). The complete expression of the field at the entrance pupil can thus be written as a linear combination of Zernike polynomials:

\begin{equation}
\begin{array}{ccl}
E_{\rm pup} & \approx & \left(1 - \dfrac{T_x^2+T_y^2}{2} \right) Z_1(\vec{r}) + i T_x Z_2(\vec{r}) + i T_y Z_3(\vec{r}) \\
& & - \dfrac{T_x^2+T_y^2}{2 \sqrt{3}} Z_4(\vec{r})  - \dfrac{2 T_x T_y}{\sqrt{6}} Z_5(\vec{r}) - \dfrac{T_x^2-T_y^2}{\sqrt{6}} Z_6(\vec{r}).
\end{array}
\end{equation}

As shown in Appendix \ref{app:small-aberration} and \ref{app:conversion-table}, the decomposition of the wavefront onto the Zernike polynomial basis under the small aberration approximation turns out to be very convenient for describing the effect of the vortex phase mask. Indeed, when propagating through the vortex focal plane mask to the Lyot plane, these Zernike modes translate into complex linear combinations of Zernike polynomials inside the geometrical pupil (the components outside the pupil are discarded here, as they are blocked by the Lyot stop).

Using the conversion table that gives the field inside the re-imaged geometrical pupil after a VC of charge $l_{\rm p}=2$ (Table \ref{tab:charge2}), we can thus directly express the electric field after the Lyot stop as

\begin{equation}
\begin{array}{cl}
E_{\rm Lyot} = & \dfrac{i T_x - T_y}{2} Z_2(\vec{r}) + \dfrac{-T_x - i T_y}{2} Z_3(\vec{r}) - \dfrac{(T_x+i T_y)^2}{2 \sqrt{3}} Z_4(\vec{r}) \\
& -i \dfrac{T_x^2+T_y^2}{2\sqrt{6}} Z_5(\vec{r}) - \dfrac{T_x^2 + T_y^2}{2 \sqrt{6}} Z_6(\vec{r}).
\end{array}
\label{eq:Elyot_ch2}
\end{equation}

\subsection{Final image analysis}

The electric field in the detector plane is obtained by the Fourier transform of Eq. \ref{eq:Elyot_ch2} (the Fourier transform of the Zernike polynomials is reminded in Eq. \ref{eq:zernike_FT})\corr{, which leads to}

\begin{equation}
\begin{array}{ccl}
E_{\rm det} & = & \dfrac{\pi}{2} (T_x + i T_y )^2 \frac{2 J_3(2 \pi \alpha)}{2 \pi \alpha} \\
&+ & \pi (T_x+i T_y) \frac{2 J_2(2 \pi \alpha)}{2 \pi \alpha} {\rm e}^{i \psi} \\
& + & \dfrac{\pi}{2} (T_x^2+T_y^2) \frac{2 J_3(2 \pi \alpha)}{2 \pi \alpha} {\rm e}^{2 i \psi},
\end{array}
\label{eq:Edet}
\end{equation}

\noindent with $(\alpha, \psi) =  \boldsymbol{\alpha}$ the polar coordinates in the image plane. The final image measured by the detector is the squared modulus of the previous expression \corr{and is thus given by}

\begin{equation}
\begin{array}{c}
I_{\rm det} =  \pi^2 \left[  \left(T_x^2+T_y^2 \right) A_2(\alpha)^2 + \frac{1}{2} \left(T_x^2+T_y^2 \right)^2 A_3(\alpha)^2  \right. \\
 + \left( (T_x^4 - T_y^4) \cos(2 \psi) + 2 (T_x^3T_y+T_y^3T_x) \sin(2 \psi) \right) A_3^2(\alpha) \\
 + \left.  2 \left( (T_x^3+T_x T_y^2) \cos(\psi) + (T_y^3+T_x^2 T_y)\sin(\psi) \right) A_2(\alpha) A_3 (\alpha)\right], \\
\end{array}
\label{eq:Idet}
\end{equation}

\begin{figure}
\centering
\begin{tabular}{c}
\includegraphics[width=.7\linewidth]{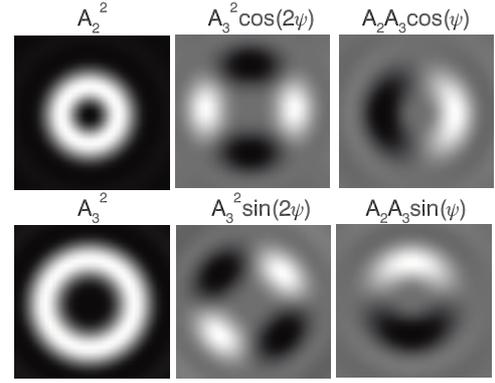} \\
\end{tabular}
\caption{Simulation of the different components that form the final image on the detector when the input wavefront is tilted, with the notations used in Eq. \ref{eq:Idet}.}
\label{fig:PSFs}
\end{figure}

\noindent with $A_i (\alpha)= \frac{2 J_i(2\pi \alpha)}{2 \pi \alpha}$. The final image consists of several terms, but only the last two contribute to the axial asymmetry in the image, as illustrated in Fig. \ref{fig:PSFs}. As a consequence, the relation between tip-tilt and the asymmetry simply writes

\begin{equation}
\begin{array}{c}
\Delta I_x =\beta  \left( T_x^3+T_x T_y^2 \right), \\
\Delta I_y =\beta  \left( T_y^3+T_y T_x^2 \right).
\end{array}
\label{eq:model}
\end{equation}

\noindent \corr{Here, $\beta$ is a normalization constant, corresponding to}

\begin{equation}
\beta =\dfrac{ 8 \pi^2 } {I_0} \int_0^{\alpha_0} A_2(\alpha) A_3(\alpha)\alpha {\rm d}\alpha,
\end{equation}

\noindent \corr{with $\alpha_0$ the maximal value allowed for $\alpha$ when integrating over the finite size quadrants (typically of few $\lambda/D$). For the sake of simplicity, the integral has been expressed using polar coordinates, but rigorously it should be rewritten in order to consider the square shape of the quadrants.}

This result is fully consistent with the model found empirically by \cite{Mas2012} for the FQPM. The only difference lies in the value of the $\alpha$ parameter of the model given by Eq.~\ref{eq:model_mas}: $\alpha_{\rm FQPM}=4$ while  $\alpha_{\rm vortex}=1$. In the vortex case, the system admits a unique solution\corr{, which can be written as}

\begin{equation}
\begin{array}{c}
T_x = \left( \dfrac{\Delta I_x}{\beta} \right)^{\frac{1}{3}} \left(\dfrac{\Delta I_x^2}{ \Delta I_x^2 + \Delta I_y^2} \right)^{\frac{1}{3}}, \\
T_y = \left( \dfrac{\Delta I_y}{\beta} \right)^{\frac{1}{3}} \left(\dfrac{\Delta I_y^2}{ \Delta I_x^2 + \Delta I_y^2} \right)^{\frac{1}{3}}.
\end{array}
\label{eq:tiptilt_estimation}
\end{equation}

\noindent The particular value of $\alpha_{\rm vortex}$ reflects the fact that a vortex phase mask is perfectly centro-symmetric, which is not the case for the FQPM. Indeed, from this system of equations, it can be shown that the simple law

\begin{equation}
\Delta I_{\theta} = \beta T_{\theta}^3
\label{eq:simple_Ix}
\end{equation}

\noindent is sufficient to describe the relation between the tip-tilt and the asymmetry, as soon as the differential intensity $\Delta I_{\theta}$ is measured along the axis of the applied tip-tilt $T_{\theta}$. In this case, the differential intensity measured in the orthogonal direction, $\Delta I_{\theta+\pi/2}$, is indeed zero. The direction of the tip-tilt can be inferred from the $\Delta I_x$ and $\Delta I_y$ measurements: $\tan\theta = \Delta I_y / \Delta I_x$, implying that $\Delta I_\theta=(\Delta I_x^2+\Delta I_y^2)^{1/2}$. In other words, it means that the cross-terms $T_xT_y^2$ and $T_yT_x^2$ in Eq.~\ref{eq:model} are not due to a cross-talk between the two axes, like in the case of the FQPM, but are rather due to a change of reference axes.

\section{Experimental validation}
\label{sec:experimental_validation}

\begin{figure}
\centering
\includegraphics[width=.9\linewidth]{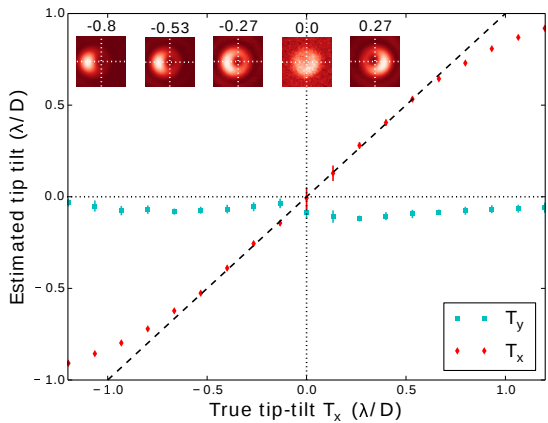}
\caption{Experimental results for the estimation of the tip-tilt aberration in one direction (the AGPM has been translated along the $x$ axis only). Error bars are computed from the standard deviation of 50 values estimated from a sequence of 50 images. The images on the top left corner of the graph show the mean images of several sequences, acquired for different values of tip-tilt that are given on top of each image in $\lambda/D$.}
\label{fig:tiptilt_results}
\end{figure}

The model describing the relation between the tip-tilt amount and the asymmetry in the VC image has been validated thanks to experimental data. Test campaigns were indeed carried out for characterizing new-generation L-band AGPMs recently manufactured at Uppsala Universitet (Vargas Catalan et al., in prep). 

These campaigns have been conducted on the YACADIRE bench at LESIA (Observatoire de Paris). This bench was used to characterize the coronagraphic masks for SPHERE \citep{Boccaletti2008} and thus mimics its optical layout ($f$/40 converging beam at the focal plane). We used a circular non-obstructed pupil and a circular Lyot stop (radius downsized by 80\% with respect to the entrance pupil radius). For the testing of the AGPMs, a cold L-band spectral filter was installed in the camera enclosure. The source is a Tungsten lamp, feeding a single-mode fibre. The bench layout is detailed in \cite{Delacroix2013}, who report on the first laboratory characterization of L-band AGPMs.

The AGPM was first centred in $x$ and $y$ with respect to the beam by minimizing the flux integrated by the camera. The position along the optical axis was also optimized. Sets of 50 images were taken for different positions of the AGPM along the $x$ axis. The $\Delta I_x$ and $\Delta I_y$ values are measured for every image. One has to note that translating the AGPM in the focal plane has not the same effect on the coronagraphic image as a tilted wavefront hitting the mask. Its shape will be affected in the same way, but it remains centred on the same position, while a tilted wavefront induces an additional translation of the image. This has been taken into account in the data processing (the quadrants were shifted by the number of pixels expected for the corresponding tip-tilt). The $T_x$ and $T_y$ are estimated for each image using Eq.~\ref{eq:tiptilt_estimation}. For one position, the final tip-tilt estimates result from the mean of the 50 estimates, and the error bar from their standard deviation. 

The results are shown in Fig. \ref{fig:tiptilt_results}. The $T_x$ estimates are in agreement with the true tip-tilt for a range of around $\pm 0.5\,\lambda/D$ from the center, where the estimations start to diverge from the expected value by more than their error bar. While $T_y=0$ was expected for the other axis, it seems that the position in that direction was not optimal and that the AGPM was probably off by about $-0.07\,\lambda/D$, corresponding to a shift of $10\,\muup$m in the focal plane. The data set has also been processed to estimate the transmission efficiency as a function of tip-tilt along the $x$ direction. These results are detailed in the Appendix \ref{app:off-axis_exp} and show that the highest extinction rate was obtained at $0.02\,\lambda/D$ from the position that was thought to be optimal during the experiment. This corresponds to a shift of $3.5\,\muup$m in the focal plane.

In conclusion, our results show that the model derived to retrieve the tip-tilt is valid for a circular non-obstructed pupil. The post-processing of the images has also shown that the manual optimization of the $x$ and $y$ position of the AGPM might not be optimal (for this particular experiment, the best manual alignment of the AGPM was off by $0.02\lambda/D$ and $0.07\lambda/D$ \corr{in x and y,} respectively), showing the limit of a manual positioning, as it is currently performed at the telescope. An automated method of tip-tilt retrieval based on the QACITS post-coronagraphic analysis will thus significantly improve the vortex phase mask centering.

\section{QACITS on a centrally obstructed pupil}
\label{sec:central_obstruction}

\begin{figure}
\centering
\begin{tabular}{l}
a) Simulated images for a circular non obstructed pupil \\
\includegraphics[width=.9\linewidth]{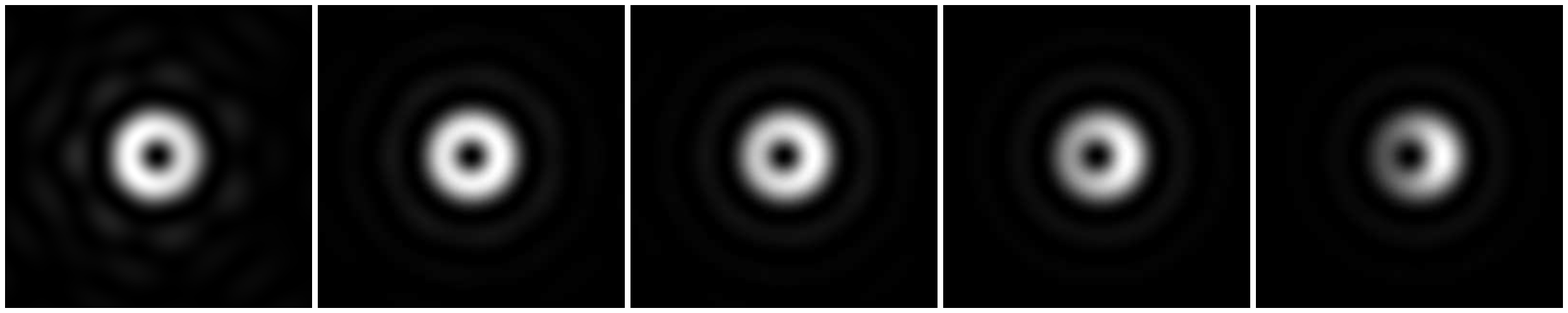} \\
\\
b) Simulated images for a circular obstructed pupil (24\%) \\
\includegraphics[width=.9\linewidth]{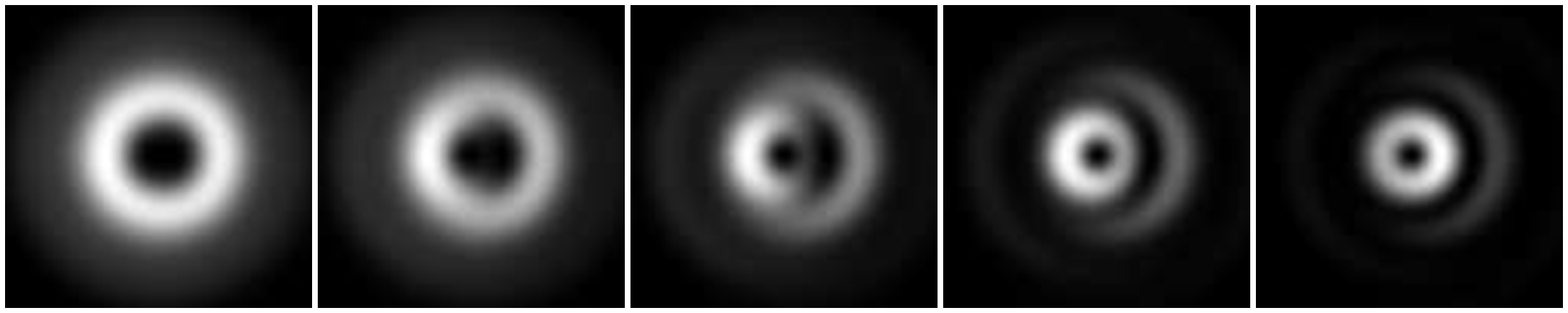} \\
\end{tabular}
\caption{Each row of images corresponds to simulated coronagraphic PSFs obtained for different tip-tilt values, from left to right: 0.01, 0.05, 0.10, 0.20 and 0.40$\,\lambda/D$. Each image intensity has been normalized by its maximum value. }
\label{fig:PSFs_obsc}
\end{figure}

All the considerations from the previous sections are valid for a circular non-obstructed pupil. However, ground-based telescopes are usually centrally obstructed by the shadow of the secondary mirror. In the case of a central obstruction, the field distribution at the Lyot plane is affected by an additional contribution that falls inside the geometrical pupil, even for an on-axis source, thus preventing from a perfect on-axis starlight rejection. This significantly impacts the final image shape, and in particular the asymmetry of the image. As illustrated in Fig. \ref{fig:PSFs_obsc}, the flux gradient changes sign for small tip-tilt in comparison with the image produced by an unobstructed pupil, implying that the model linking the differential intensity and the tip-tilt is different and more complex. In the following section, we analyse the theoretical model and adapt our QACITS tip-tilt estimator.

\subsection{Analytical derivation of the model}

Following the superposition principle, the entrance pupil can be written down as the sum of a positive contribution for the circular non-obstructed pupil and a negative contribution for the central obstruction of radius $\tau < 1$ (the pupil is defined with a radius of 1 when using Zernike polynomials)\corr{, that is}

\begin{equation}
E_{\rm pup}^{\rm } = E_{\rm pup}^0 + E_{\rm pup}^{\rm obsc} = {\rm e}^{i \phi_0} - {\rm e}^{i \phi_{\rm obsc}},
\end{equation}

\noindent with $\phi_0=\phi$ as defined in Eq. \ref{eq:phase}. \corr{The phase term}

\begin{equation}
\Phi_{\rm obsc} = \tau \left[ T_x Z_2 \left( \frac{r}{\tau} \right) + T_y Z_3 \left( \frac{r}{\tau} \right) \right]
\end{equation}

\noindent defines the phase of the component inside the central obstruction, using scaled Zernike polynomials (defined for $r/\tau<1$), such that the total field in the central obstructed area is cancelled out. As a consequence, the electric field at the Lyot plane (after the Lyot stop) will be composed of all the terms already mentioned in Eq. \ref{eq:Elyot_ch2} and of the additional following terms arising from the presence of the central obstruction:

\begin{equation}
\begin{array}{cc}
E_{\rm Lyot}^{\rm obsc} &= \tau^2 \left( \tau (T_x^2 + T_y^2)- 1 \right) \dfrac{{\rm e}^{2 i \theta}}{r^2} - \tau^4 \left( T_y + i T_x \right) \dfrac{{\rm e}^{3 i \theta}}{r^3} \\
& + \tau ^6 \left( \dfrac{T_x^2-T_y^2}{4} -i T_x T_y \right) \dfrac{{\rm e}^{4 i \theta}}{r^4}.
\end{array}
\label{eq:Elyot_obsc}
\end{equation}

\noindent Basically, these terms correspond to the decaying exponential terms that appear outside the geometrical pupil (see Appendix \ref{app:conversion-table} and in particular Eq.~\ref{eq:Tnk}) of radius $\rho$, since the components inside the obstruction are blocked by the Lyot stop. As a consequence, these terms are defined for $r > \tau$ and $r < 1$ (inner and outer diameter of the Lyot stop). The \corr{Fourier transform of a} function of the general form ${\rm e}^{i k \theta}/r^k$, restrained to this domain \corr{can be written as}

\begin{equation}
\mathcal{F}\left[ \dfrac{{\rm e}^{i k \theta}}{r^k} \right] =\pi i^k {\rm e}^{i k \psi} \left[ A_{k-1}(\alpha) - A_{k-1}(\alpha \tau) \right],
\end{equation}

\noindent \corr{so that} the electric field on the detector due to the central obstruction can be written as

\begin{equation}
\begin{array}{cccl}
E_{\rm det}^{\rm obsc} &=&\pi  \left( 1- \tau (T_x^2+T_y^2) \right) {\rm e}^{2 i \psi} & \tau^2  \Delta A_1(\alpha)\\
&+ &\pi (i T_y - T_x){\rm e}^{3 i \psi} & \tau^4 \Delta A_2(\alpha) \\
&  + &\pi \frac{T_x^2-T_y^2-4iT_xT_y}{4} {\rm e}^{4 i \psi}  &  \tau^6 \Delta A_3(\alpha),
\end{array}
\label{eq:Edet_obsc}
\end{equation}

\noindent with $\Delta A_k(\alpha)=A_k(\alpha)-A_k(\alpha \tau)/\tau^{k-1}$. Numerical estimations show that the $\Delta A_1 \tau^{2}$ component is the dominant term, the two other ones \corr{being} significantly \corr{smaller} in absolute values due to the factor $\tau^{2k}$ (because $\tau <1$). Since it would be uselessly painful to derive the complete expression of the intensity recorded by the detector, we choose to neglect the two weaker terms in the following computation. In addition, we can approximate the factor $1-\tau(T_x^2+T_y^2) \approx 1$, thus assuming that the tip-tilt has a negligible effect on the light diffracted by the central obstruction. As a consequence, the intensity on the detector consists of the expression given in Eq.~\ref{eq:Idet} augmented by the following terms (calculated from the modulus of the first term of Eq.\,\ref{eq:Edet_obsc} and cross-terms between the terms of Eq.\,\ref{eq:Edet} and first term of Eq.\,\ref{eq:Edet_obsc}):

\begin{equation}
\begin{array}{rl}
I_{\rm det}^{\rm obsc} =& \pi^2 \tau^2 \left[ ~~ \tau^2 \times\Delta A_1 (\alpha)^2 \right. \\
+ &  (T_x^2+T_y^2) \times \Delta A_1 (\alpha)A_3(\alpha) \\
+ &  \left( (T_x^2-T_y^2) \cos(2\psi) +2T_xT_y \sin(2\psi) \right) \times \Delta A_1 (\alpha) A_3(\alpha) \\
+ & \left. \left( 2 T_x\cos(\psi) +2T_y \sin(\psi) \right) \times \Delta A_1(\alpha) A_2(\alpha) ~~ \right]. \\
\end{array}
\label{eq:Idet_obsc}
\end{equation}

\begin{figure}
\centering
\includegraphics[width=.75\linewidth]{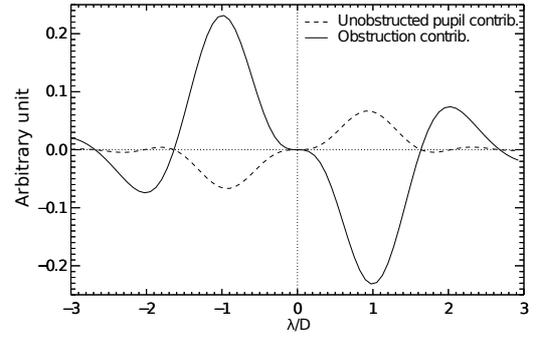}
\caption{Horizontal profile ($\psi=0$) for the two asymmetric contributions in the final image plane, due to the whole pupil (\corr{i.e.} $A_2(\alpha) A_3(\alpha)  $) and to the central obstruction (\corr{i.e.} $\Delta A_1(\alpha) A_2(\alpha) $).}
\label{fig:asym_prof}
\end{figure}

\noindent Only the last term \corr{of Eq.\,\ref{eq:Idet_obsc}} produces an asymmetric pattern with respect to the $x$ and $y$ axes. It has to be noted that the principal lobe of $\Delta A_1(\alpha) A_2(\alpha)$ has negative values, so that this term will be in competition with the asymmetric term arising from the circular unobstructed pupil (last term of Eq.~\ref{eq:Idet}). This is illustrated in Fig.~\ref{fig:asym_prof}, showing the horizontal profiles for each contribution. In addition, the contribution of the central obstruction is weighted by a coefficient directly proportional to the amount of tip-tilt, while the contribution of the circular pupil is lessen by the cube of the amount of tip-tilt. This explains why, for very small tip-tilt, the asymmetry in the images simulated with an annular pupil appears with a gradient of opposite sign compared with images simulated for an unobstructed pupil (Fig.~\ref{fig:PSFs_obsc}). Therefore, the relation between the tip-tilt and the asymmetry in the image can be written as

\begin{equation}
\begin{array}{c}
\Delta I_x = \beta \left(T_x^3 + T_x T_y^2 \right) + \gamma T_x, \\
\Delta I_y = \beta \left(T_y^3 + T_y T_x^2 \right) + \gamma T_y,
\end{array}
\label{eq:DeltaI_obsc}
\end{equation} 

\noindent with $\beta$ and $\gamma$ two real parameters of opposite signs. As it will be illustrated in the following section, the main issue with this model is that it will necessarily limit the range where the standard QACITS method can be applied, because the competition between the two terms leads to a possible ambiguity to retrieve the tip-tilt from a single intensity measurement. It also reduces the sensitivity, as the two contributions partially cancel each other. That is why a dual area QACITS method is proposed in the next section.

\subsection{QACITS in dual areas}

\begin{figure}
\centering
\begin{tabular}{cc}
\multicolumn{2}{l}{a) Standard, inner and outer areas} \\
\multicolumn{2}{r}{\includegraphics[width=.80\linewidth]{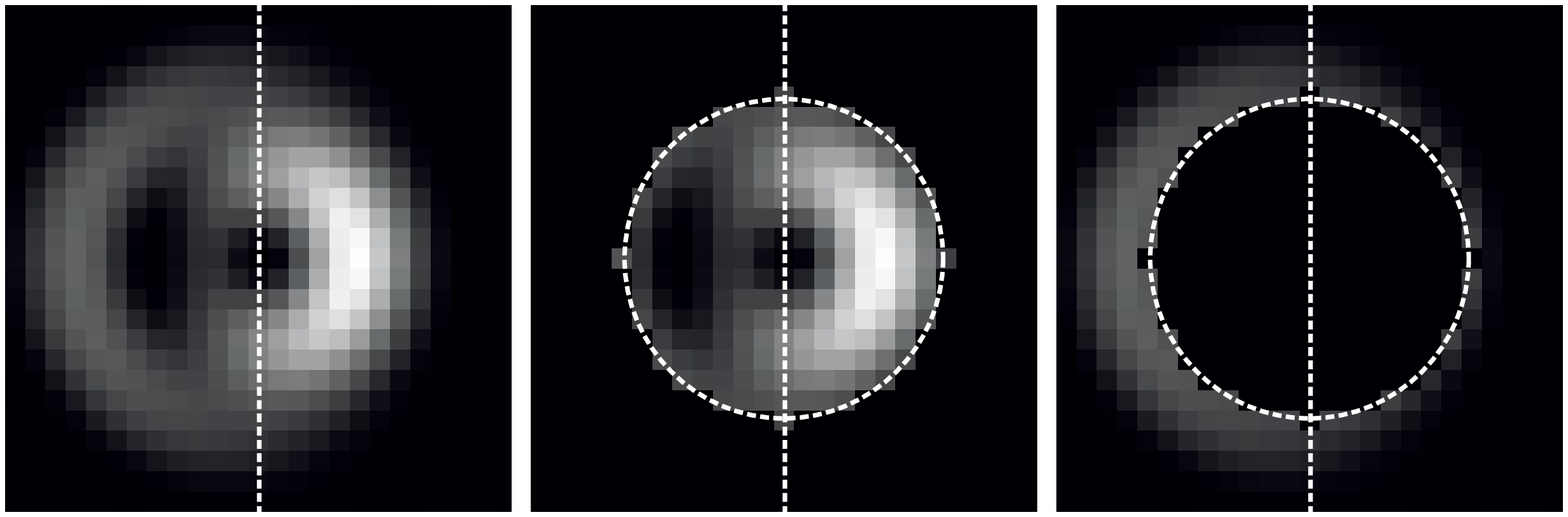}} \\ \\
\multicolumn{2}{l}{b) Flux repartition between the inner and outer areas} \\
~~~~~ Unobstructed pupil & ~~~~~ Obstructed pupil\\
\includegraphics[width=.4\linewidth]{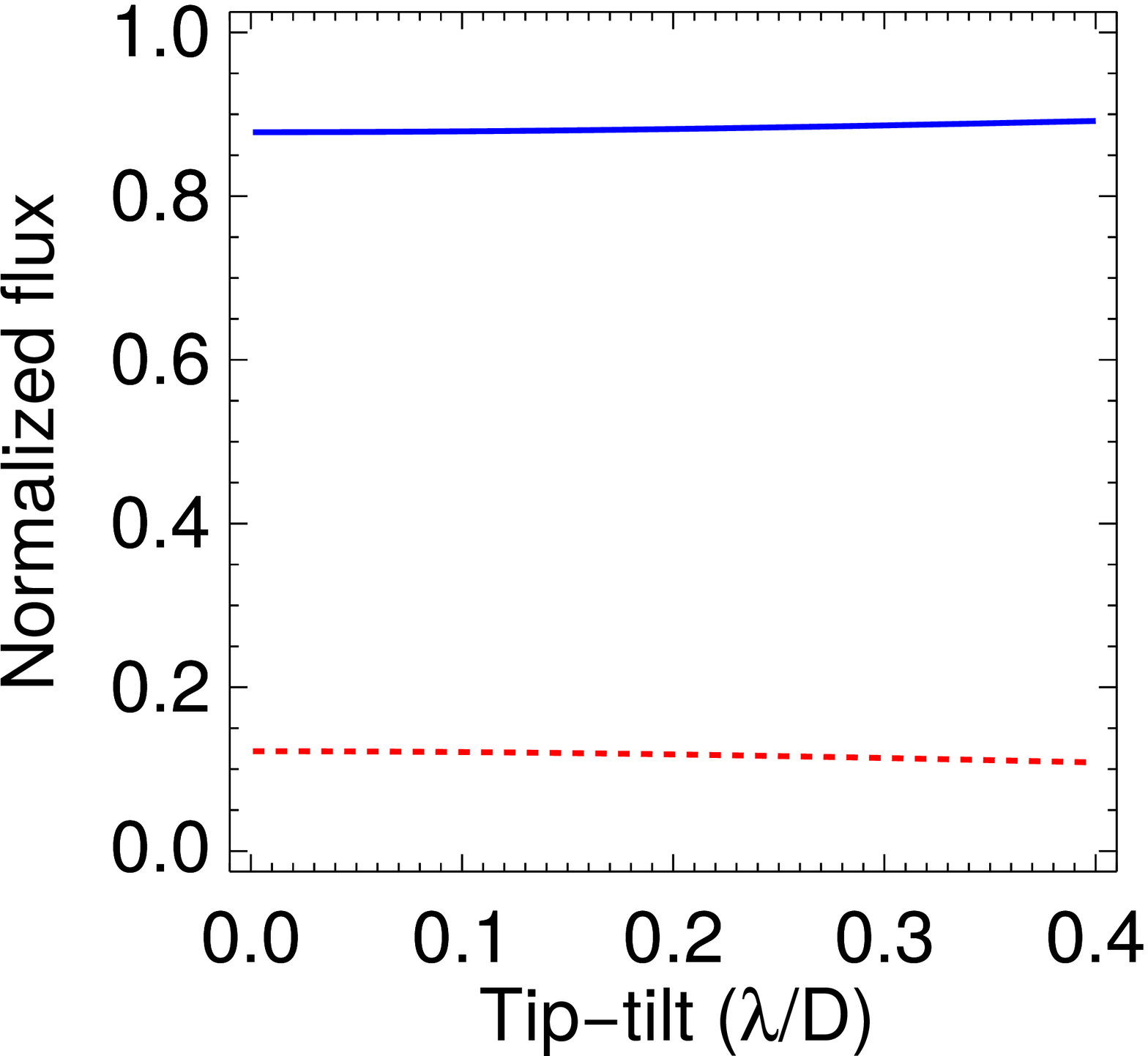} &
\includegraphics[width=.4\linewidth]{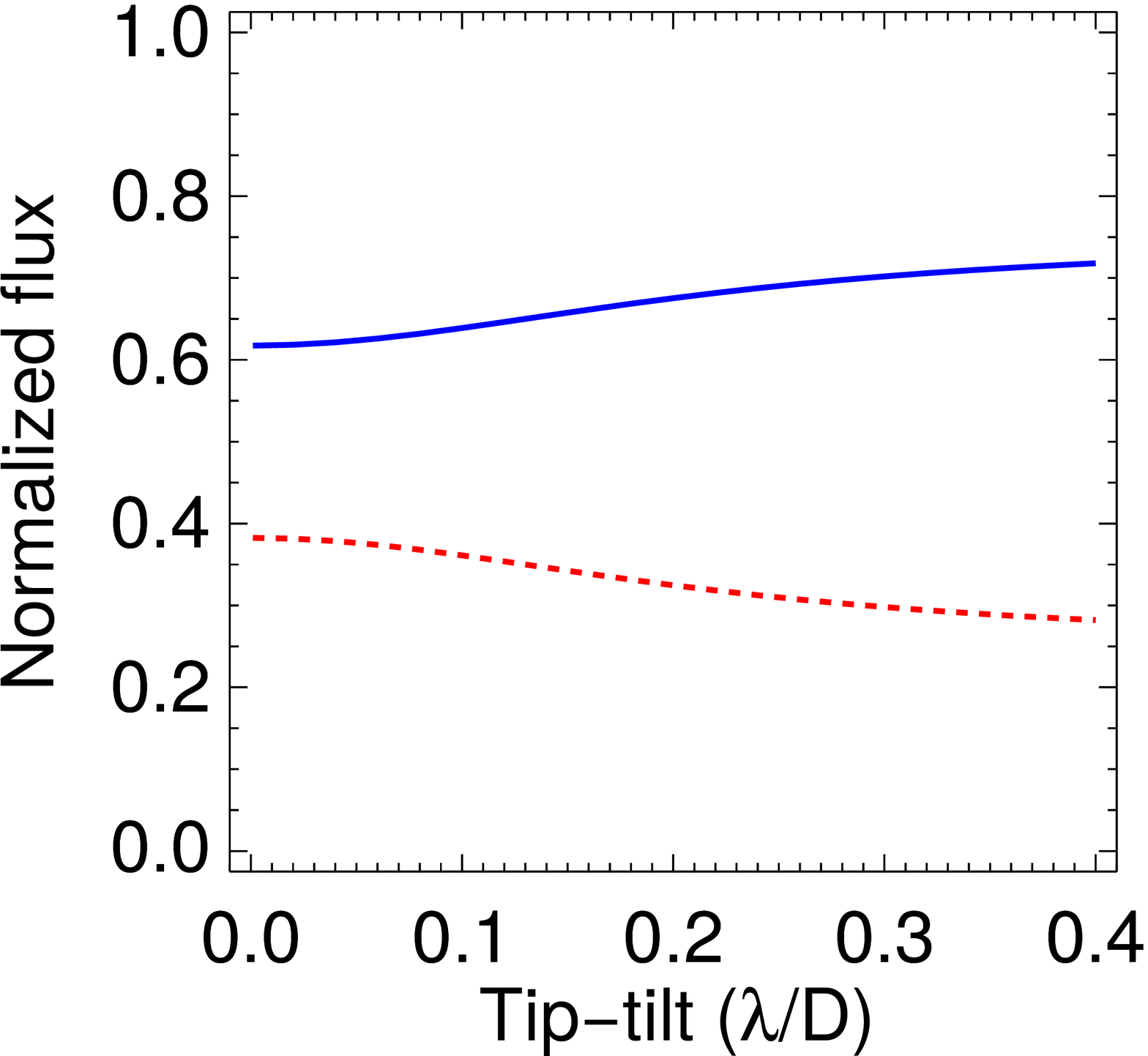}
\end{tabular}
\caption{a) Simulated image for a centrally obstructed pupil (24\% of the diameter pupil) with $0.2\,\lambda/D$ tip-tilt, showing the spatial flux distribution for the standard, inner and outer areas. The total image width is $6\,\lambda/D$ while the circle has a radius of $2\,\lambda/D$. b) The flux repartition between the inner (blue continuous line) and the outer (red dashed line) areas as a function of tip-tilt (estimated from simulated images).}
\label{fig:inner-outer}
\end{figure}

As illustrated in Fig. \ref{fig:asym_prof}, the asymmetric contribution due to the central obstruction undergoes a sign inversion at $1.6\,\lambda/D$, which corresponds to the zero of the $A_2(\alpha)$ function. 
Two areas can thus be defined in the image: the inner area ($r<1.6\,\lambda/D$) and the outer area ($r>1.6\,\lambda/D$). Note that when the outer diameter of the Lyot stop is downsized, this $1.6\,\lambda/D$ boundary has to be scaled proportionally (for instance, a Lyot stop downsized by 80\% has for effect to push the boundary to $2\,\lambda/D$). These areas are shown on a simulated image in Fig.\,\ref{fig:inner-outer}a, highlighting the fact that the intensity gradient has opposite sign depending on the considered region. Fig.\,\ref{fig:inner-outer}b shows the flux repartition between these areas for an unobstructed circular pupil and an obstructed pupil. While in the ideal unobstructed case, the flux is mostly concentrated in the central lobe ($\sim80\%$ of the total flux), a significant portion of the flux spreads outwards in presence of a central obstruction, making this dual measurement legit.

\begin{figure}
\centering
\includegraphics[width=.9\linewidth]{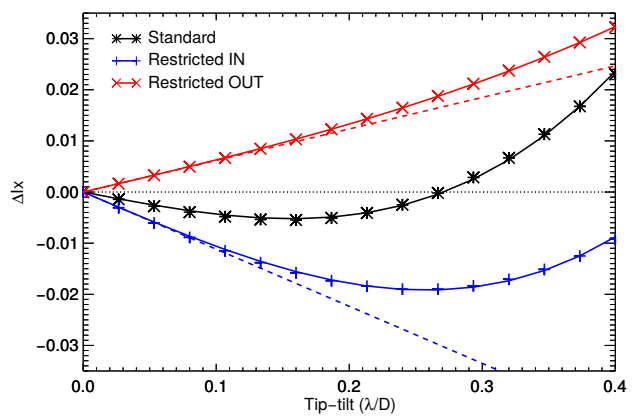}
\caption{Estimated differential intensities resulting from simulated images in the case of a pupil centrally obstructed (24\% of the full diameter). The solid line curves show the best fit model in the least squares sense (the model consists of a linear and a cubic component), while the dashed lines show the linear contribution only. The three cases differ in the area used to compute the differential intensity: standard whole area (in black), inner ($<2\lambda/D$, in blue) or outer area ($>2 \lambda/D$, in red). The outer diameter of the Lyot stop is downsized by a factor of 80\%, while the central obstruction diameter is set to 35\% (1.45 oversizing factor). }
\label{fig:Ixvstt_obsc}
\end{figure}

The differential intensities corresponding to the standard QACITS, and to the QACITS split down into inner and outer areas are shown in Fig. \ref{fig:Ixvstt_obsc} for the case of an annular pupil (24\% obstruction in diameter). While the differential intensities computed in the standard way show a degeneracy and a limited amplitude, the intensities restricted to the inner and outer areas reach higher absolute values. An interesting feature appearing in these plots is the fact that for small tip-tilt ($<0.2\lambda/D$), the model can be approximated by the linear part of the model, which dominates over the cubic term. For the sake of simplicity, we will use this approximation thereafter, especially since the system of equations (Eq.\,\ref{eq:DeltaI_obsc}) does not lead to simple analytical solutions.

\begin{table}
\caption{$\beta$ and $\gamma$ parameters estimated from simulations and based on the whole central image (standard method) or only on the inner or outer areas of the image. Different Lyot stop configurations have been simulated, with the first line corresponding to an unobstructed entrance pupil (hence the non specified inner diameter of the Lyot stop), while all the other cases result from an annular entrance pupil with 24\% central obstruction.}
\label{tab:beta_gamma}
\centering
\begin{tabular}{cc|cccccc}
\hline \hline
$L_{\rm out}$ & $L_{\rm in}$ & \multicolumn{2}{c}{Stand.}& \multicolumn{2}{c}{In. area}&\multicolumn{2}{c}{Out. area}\\
(\%) & (\%) & $\beta$ & $\gamma$ & $\beta_{\rm in}$ & $\gamma_{\rm in}$ & $\beta_{\rm out}$ & $\gamma_{\rm out}$ \\
\hline
100	&   /	& 1.08 & & 0.94 & & 0.07 & \\
100	& 24	& 0.93& -0.04& 0.88&-0.10&0.04&0.06\\
100	& 35	& 0.90 & -0.04 & 0.79 & -0.10& 0.11 & 0.05 \\
80 	& 24 & 0.75 & -0.05 & 0.70 & -0.13 & 0.05 & 0.08 \\
80	& 35	& 0.68 & -0.05 & 0.56 & -0.11&0.12&0.06\\
\hline
\end{tabular}
\end{table}

The  $\beta$ and $\gamma$ parameters defining the model given in Eq. \ref{eq:DeltaI_obsc} have been estimated thanks to simulations for different pupil configurations, and in particular different Lyot stop parameter values: the inner and outer diameter, $L_{\rm in}$ and $L_{\rm out}$, defined as a fraction of the entrance pupil diameter $D$. The $\beta$ and $\gamma$ parameters correspond to the cubic and linear components respectively and are computed by fitting the simulated points in the least-squares sense. The values are reported in Table \ref{tab:beta_gamma}. These results show that the $\gamma$ parameter weighing the linear part of the model increases with the reduction of the outer diameter of the Lyot stop mask. This is expected since the flux due to the diffraction by the central obstruction mainly distributes to the area close to the central obstruction (see the decaying exponential functions of Eq.\,\ref{eq:Elyot_obsc}), while the tip-tilt energy coming from the whole pupil is spread over the whole pupil. As a consequence, cropping part of the outer rim of the pupil implies that the central obstruction contribution, which is the source of the linear dependency, becomes relatively stronger.

\begin{figure}
\centering
\includegraphics[width=.9\linewidth]{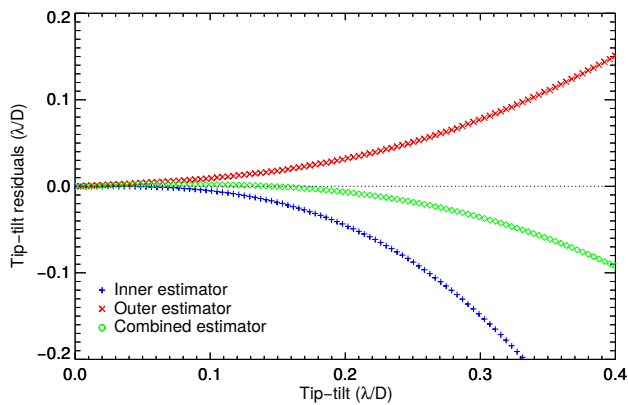}
\caption{\corr{Tip-tilt residuals obtained for the same simulation parameters as Fig.~\ref{fig:Ixvstt_obsc} (i.e. 24\% central obstruction, Lyot stop of 35\% and 80\% inner and outer diameter).}}
\label{fig:est-vs-true}
\end{figure}

The parameters have also been estimated for measurements restricted to the inner and outer areas. In both cases, the $\gamma$ parameter reaches higher values, and thus provides a better dynamic, in comparison with the values obtained by integrating the flux in the whole image (standard method). \corr{The final estimator is therefore taken as the average of the inner and outer estimators based on the linear approximation of the model, and can thus be written as
\begin{equation}
\boldsymbol{T}^{\rm est} = \frac{1}{2} \left( \frac{\boldsymbol{\Delta I} ^{\rm in}}{ \gamma_{\rm in}} +\frac{ \boldsymbol{\Delta I} ^ {\rm out}}{ \gamma_{\rm out}} \right),
\end{equation}
\noindent with $\boldsymbol{T}^{\rm est}$ and $\boldsymbol{\Delta I}$ defined as vectors with the $x$ and $y$ components of the tip-tilt estimate and differential intensity measurements respectively. The exponents "in" and "out" refer to the area of the image used to integrate the flux, namely inner or outer part. This average estimator has been applied to simulated images affected by a tip-tilt ranging from $0$ to $0.4\,\lambda/D$. The tip-tilt residuals are reported in Fig.~\ref{fig:est-vs-true}. The mismatch between the model and the linear approximation induces a bias in the inner and outer estimators. These biases happen to be of opposite sign, and thus compensate each other at least partially when taking the average. In practice, this offset is not critical since the QACITS algorithm is supposed to be used in closed loop control. The results show that for large tip-tilt amounts, the combined estimator under-estimates the amplitude, which means that the convergence might be slower at first.}

To conclude, we have derived the theoretical model and modified the QACITS estimator to make it applicable to the case of a centrally obstructed aperture. Because the contribution of the obstruction counterbalances the contribution of the circular pupil, the validity range is reduced to small tip-tilt amounts (\corr{for tip-tilt $<0.2 \lambda/D$, the bias is smaller than 3\%}). However, the presence of the central obstruction is responsible for a higher starlight leakage (at least $5\%$ for a central obstruction of $24\%$ in diameter), providing a better sensitivity, but also a better dynamic due to the linearity of the model, as opposed to the cubic model in the non obstructed case.

\section{Performance in presence of higher order aberrations}
\label{sec:aberrations}

\begin{figure*}
\centering
\begin{tabular}{ll}
\includegraphics[width=.4\linewidth]{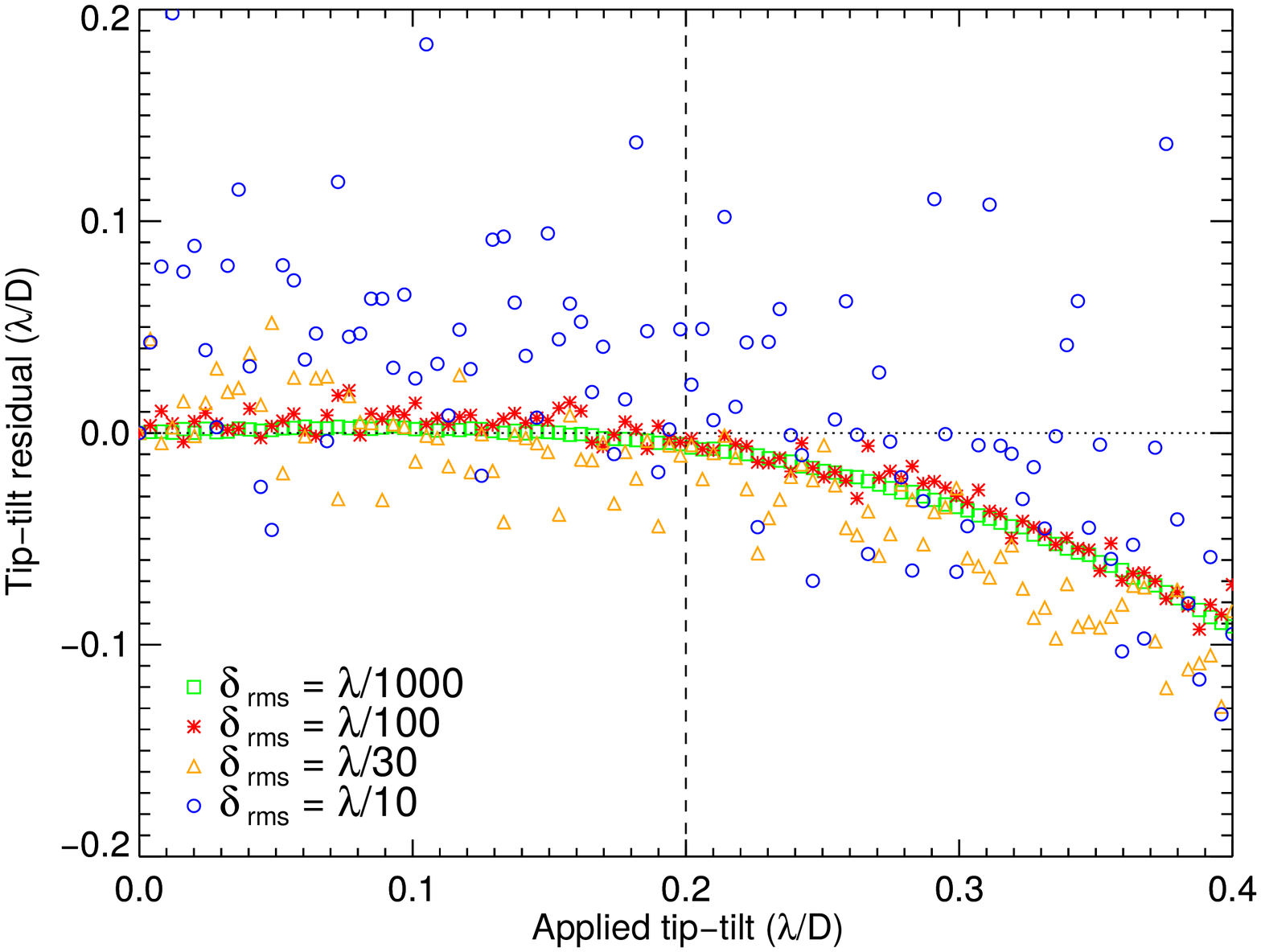} & \includegraphics[width=.4\linewidth]{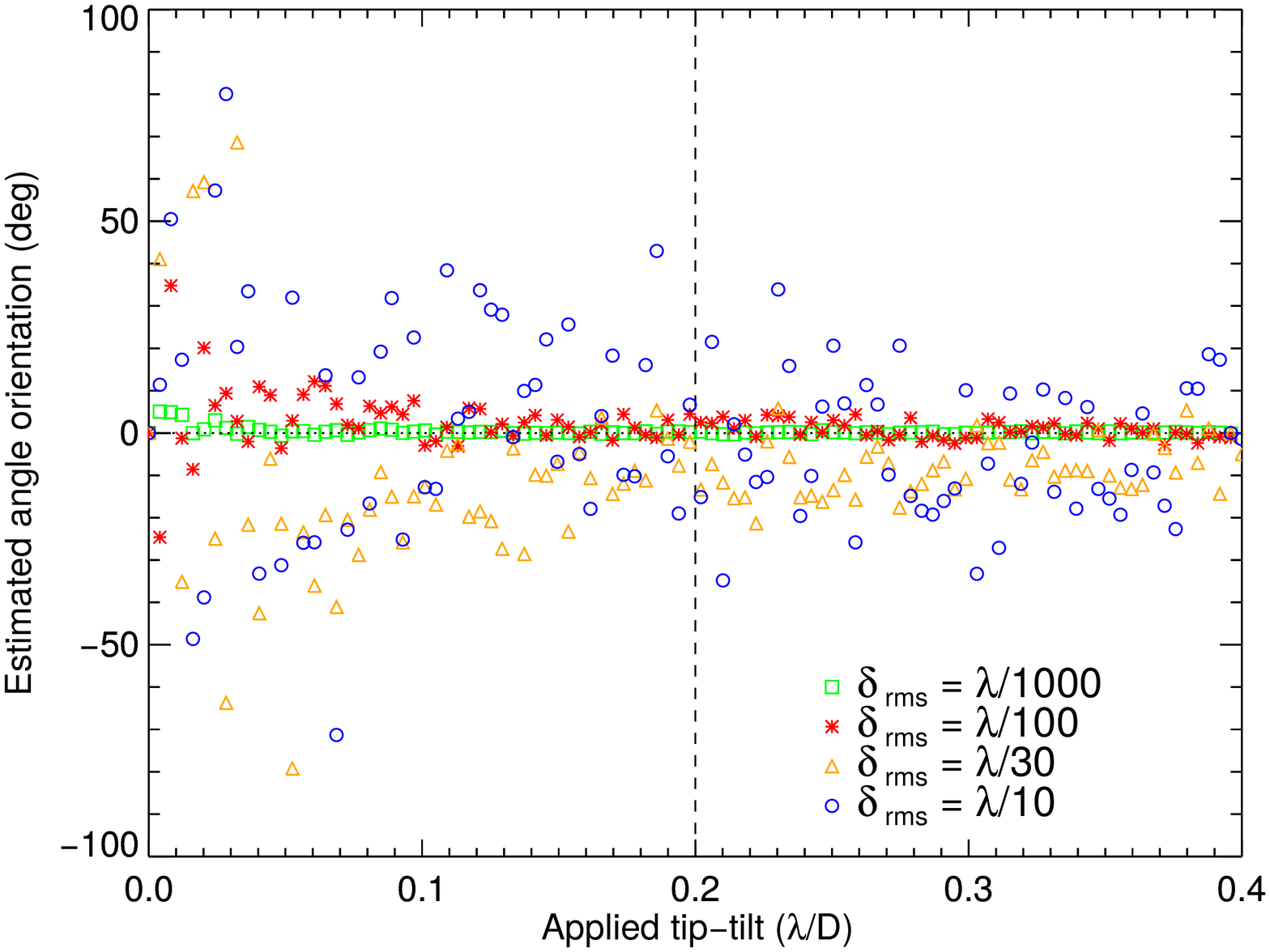} 
\end{tabular}
\caption{Simulation results of tip-tilt estimation using QACITS with a centrally obstructed pupil in presence of higher order aberrations. Amplitude residuals are shown on the left, while orientation angle residuals are shown on the right, for different levels of aberrations.}
\label{fig:aberr_simu_points}
\end{figure*}

\begin{figure*}
\centering
\begin{tabular}{ll}
\includegraphics[width=.4\linewidth]{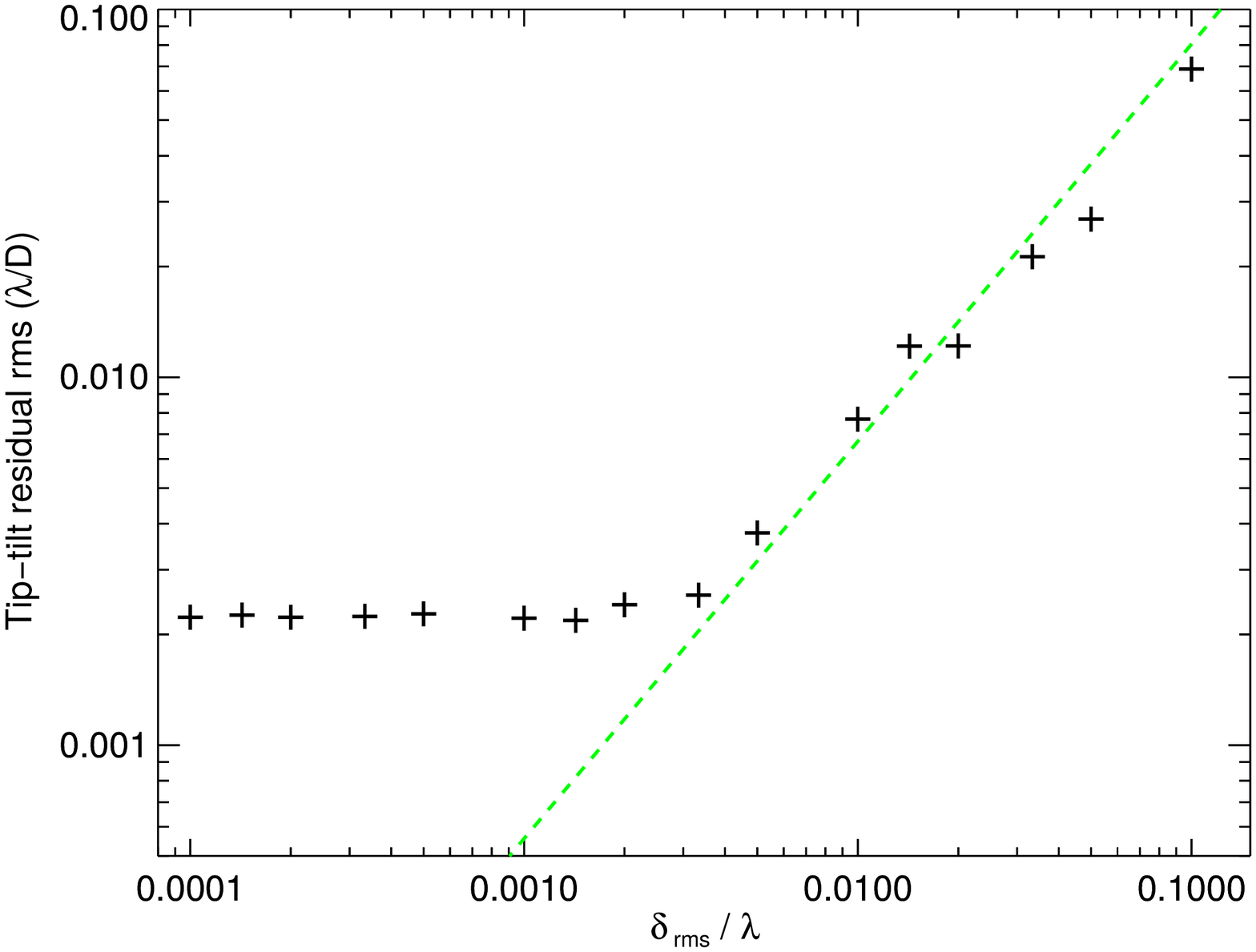} & \includegraphics[width=.4\linewidth]{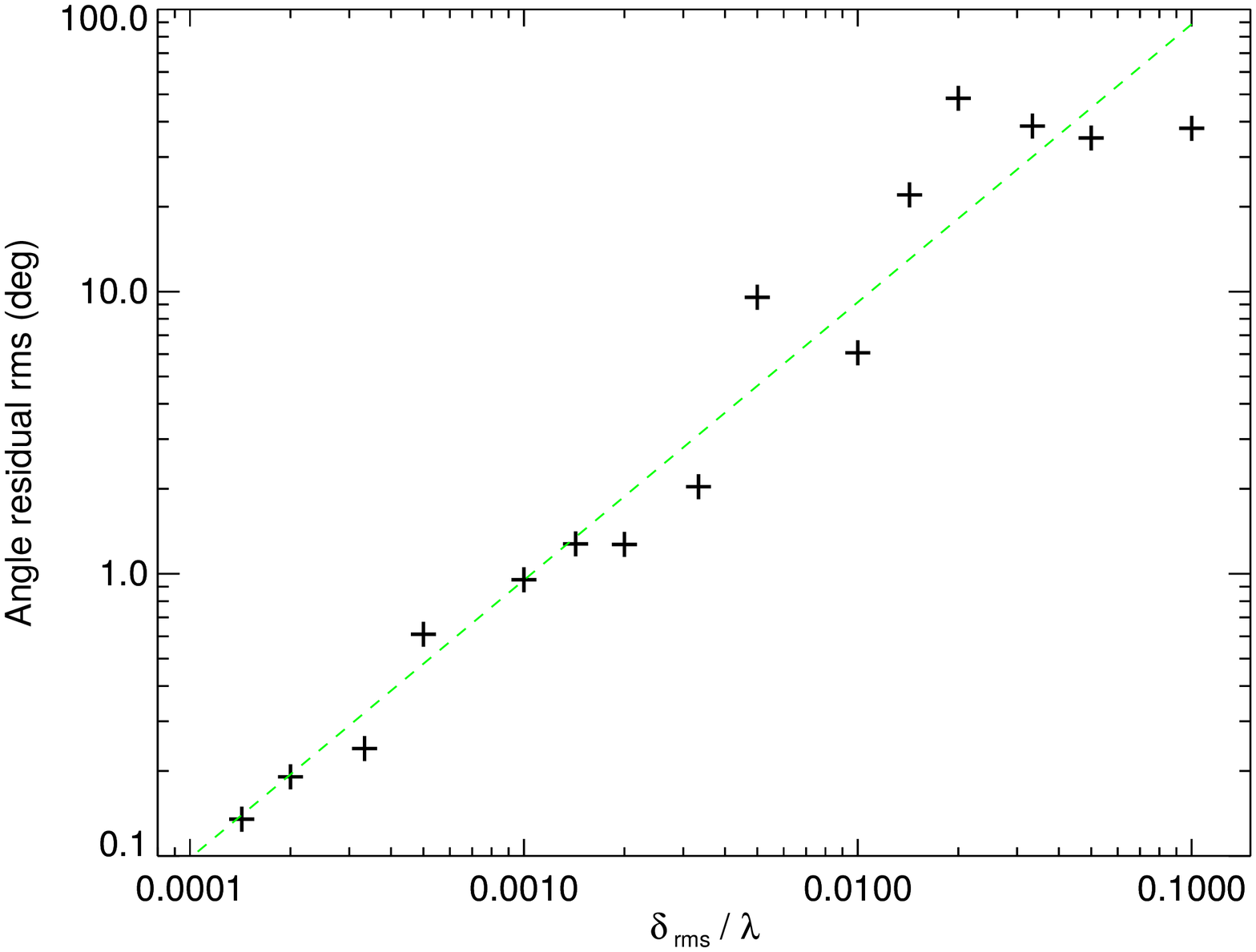} \\
\end{tabular}
\caption{Root mean square values for the residual tip-tilt amplitude (left) and orientation angle (right), as a function of the wavefront error. Note that for the amplitude, the rms is computed over the reduced [0,0.2]$\lambda/D$ range, where the linear model approximation is valid. The green dashed lines correspond to a best fit power law model.}
\label{fig:aberr_simu_rms}
\end{figure*}

In practice, real wavefronts are affected not only by tip-tilt but also by higher order aberrations. Static aberrations due to imperfect optics surfaces can be handled by subtracting a reference image. However, quasi-static speckle patterns may corrupt the $\Delta I_x$ and $\Delta I_y$ measurements. Such aberrations may be caused by temperature and mechanical drifts, that slowly evolve with time, and are not sensed by the adaptive optics system. In order to quantify the effect of higher order aberrations, simulations were conducted with phase screens generated from a power spectral density defined as the inverse power law of exponent~$2$. This kind of model is typical for fractal finish surface quality \citep{Church1988}, like the high quality optics of the SPHERE instrument \citep{Dohlen2011}. The simulated coronagraph is based on a circular entrance pupil obstructed by 24\% of its diameter and a Lyot stop with an oversized central obstruction of 35\% and outer diameter of 80\% of the initial entrance pupil (this corresponds to the typical obstruction and Lyot configuration of the \corr{NIRC2 instrument at the} Keck telescope).

Every phase screen is drawn randomly. The tip-tilt component is estimated by a projection onto the base of Zernike polynomials and subtracted. A hundred tip-tilt values ranging from 0 to 0.4\,$\lambda/D$ are uniformly drawn and applied to the wavefront in the horizontal direction (orientation angle $\theta = 0$). The tip-tilt amplitude and orientation angle are then estimated from the image with the dual QACITS method. The final estimate is computed from the average of both estimators using the inner (<2\,$\lambda/D$) and the outer part of the image (2\,$\lambda/D$<$\alpha$<3\,$\lambda/D$). As shown in the previous section, for small tip-tilt values (<$0.2\,\lambda/D$), the relation between the asymmetry in the image and the tip-tilt amount can be approximated by a linear function, whose proportionality factors, $\gamma$, are given in Table \ref{tab:beta_gamma}, i.e.  $\gamma_{\rm in}=-0.11$ and $\gamma_{\rm out}=0.06$ in the configuration used in our simulations.

The results are shown in Fig.\,\ref{fig:aberr_simu_points}. Different aberration levels have been simulated, from $\delta_{rms}=\lambda/10^4$ to $\delta_{rms}=\lambda/10$. As expected, for large tip-tilt amounts (>0.2\,$\lambda/D$), the model is not valid any more and a bias appears, in the amplitude estimation in particular. 
The root mean square (rms) values of the tip-tilt amplitude residuals reported in Fig.~\ref{fig:aberr_simu_rms} \corr{have therefore been computed on the reduced range of tip-tilt $<0.2\lambda/D$. These results} show that in the small tip-tilt regime and for very low aberration levels ($\delta_{rms}<\lambda/300$), the \corr{bias due to the linear approximation} dominates the speckle noise, and limits the accuracy of the estimation to $2.2 \times  10^{-3}\,\lambda/D$. For higher aberration levels, the \corr{accuracy is} dominated by the effect of the aberration, and the tip-tilt residual rms increases linearly with the wavefront error rms, expressed as a fraction of wavelength. This is observed for the amplitude  as well as for the orientation angle (the slope of the best fit models drawn in Fig.~\ref{fig:aberr_simu_rms} in log-log scale is 1.0 for both cases).

These results illustrate the stability level we can expect from a control loop based on the QACITS technique when higher order aberrations affect the PSF shape. The tip-tilt affecting the beam can be estimated with a precision better than $10^{-2}\,\lambda/D$ and $5 \times 10^{-2}\,\lambda/D$ in presence of wavefront errors up to $\delta_{\rm rms}=\lambda/70$ and $\delta_{\rm rms}=\lambda/14$ respectively (corresponding to $\sim 50$ nm rms and $\sim 270$ nm rms at 3.75$\muup$m). However, quasi-static speckles tend to evolve slowly with time (i.e. on minute time scales). Therefore, in practice, two consecutive images are not completely decorrelated, unlike our set of simulated phase screens, and in this case part of the high order aberration impact can be avoided by subtracting a reference image, obtained for the best centering of the coronagraphic mask.

\section{Conclusions and prospects}

We have described the QACITS technique for the vortex coronagraph, a method originally introduced in the case of the four-quadrant phase mask \citep{Mas2012} and a circular non obstructed aperture. We have derived the analytical model for the VC and found a cubic power law, validated by simulations and experimental results. However, the presence of a central circular obstruction adds a linear component that induces an intensity gradient in the opposite direction. In order to tackle this more complex model, we have introduced the QACITS method in dual zones (distinguishing the inner lobe from the external region), which allows the disentanglement of the cubic and linear components.

Simulations of a typical telescope configuration carried out in the presence of higher order aberrations show that the QACITS method provides an estimation of the tip-tilt with a precision of $10^{-2}\,\lambda/D$ for wavefront errors amounting to $\lambda/70$ rms. For very low level of aberrations ($<\lambda/300$), systematic errors arising from the linear approximation of the model limit the accuracy of the estimation to $2.2 \times 10^{-3}\,\lambda/D$. The practical implementation may also be limited by other factors, such as the brightness of the star, or the possible asymmetry of the observed object. This aspect will be discussed in more detail in another paper.

It can also be emphasized that the Zernike-based analysis reported in the Appendices highlights a remarkable feature of the vortex coronagraph: at first order, small aberrations expressed as Zernike polynomials simply translate into a complex linear combination of other Zernike polynomials in the Lyot plane. We are currently investigating other wavefront sensing techniques exploiting this characteristic.

To conclude, the QACITS technique offers an easy way to control the centering of the vortex phase mask directly from the scientific image, thus avoiding non-common path errors that an additional wavefront sensor fails to measure. Its simplicity of implementation makes QACITS a valuable and directly available tool for all the instruments equipped with a vortex phase mask.

\corr{
\begin{acknowledgements}
The research leading to these results has received funding from the European Research Council under the European Union's Seventh Framework Programme (ERC Grant Agreement n. 337569) and from the French Community of Belgium through an ARC grant for Concerted Research Action.
\end{acknowledgements}
}

\bibliographystyle{aa}
\bibliography{qacits_biblio}

\begin{thebibliography}{29}
\expandafter\ifx\csname natexlab\endcsname\relax\def\natexlab#1{#1}\fi

\bibitem[{{Absil} {et~al.}(2013){Absil}, {Milli}, {Mawet}, {Lagrange},
  {Girard}, {Chauvin}, {Boccaletti}, {Delacroix}, \& {Surdej}}]{Absil2013}
{Absil}, O., {Milli}, J., {Mawet}, D., {et~al.} 2013, \aap, 559, L12

\bibitem[{{Baudoz} {et~al.}(2010){Baudoz}, {Dorn}, {Lizon}, {Fusco}, {Dohlen},
  {Charton}, {Beuzit}, {Puget}, {Mouillet}, {Felt}, {Wildi}, {Barrufolo},
  {Kasper}, \& {Hubin}}]{Baudoz2010}
{Baudoz}, P., {Dorn}, R.~J., {Lizon}, J.-L., {et~al.} 2010, \procspie, 7735, 5

\bibitem[{{Boccaletti} {et~al.}(2008){Boccaletti}, {Abe}, {Baudrand}, {Daban},
  {Douet}, {Guerri}, {Robbe-Dubois}, {Bendjoya}, {Dohlen}, \&
  {Mawet}}]{Boccaletti2008}
{Boccaletti}, A., {Abe}, L., {Baudrand}, J., {et~al.} 2008, \procspie, 7015, 1

\bibitem[{{Brandl} {et~al.}(2014){Brandl}, {Feldt}, {Glasse}, {Guedel},
  {Heikamp}, {Kenworthy}, {Lenzen}, {Meyer}, {Molster}, {Paalvast}, {Pantin},
  {Quanz}, {Schmalzl}, {Stuik}, {Venema}, \& {Waelkens}}]{Brandl2014}
{Brandl}, B.~R., {Feldt}, M., {Glasse}, A., {et~al.} 2014, \procspie, 9147, 21

\bibitem[{Church(1988)}]{Church1988}
Church, E.~L. 1988, Applied Optics, 27, 1518

\bibitem[{{Defr{\`e}re} {et~al.}(2014){Defr{\`e}re}, {Absil}, {Hinz}, {Kuhn},
  {Mawet}, {Mennesson}, {Skemer}, {Wallace}, {Bailey}, {Downey}, {Delacroix},
  {Durney}, {Forsberg}, {Gomez}, {Habraken}, {Hoffmann}, {Karlsson},
  {Kenworthy}, {Leisenring}, {Montoya}, {Pueyo}, {Skrutskie}, \&
  {Surdej}}]{Defrere2014}
{Defr{\`e}re}, D., {Absil}, O., {Hinz}, P., {et~al.} 2014, \procspie, 9148, 3

\bibitem[{{Delacroix} {et~al.}(2013){Delacroix}, {Absil}, {Forsberg}, {Mawet},
  {Christiaens}, {Karlsson}, {Boccaletti}, {Baudoz}, {Kuittinen}, {Vartiainen},
  {Surdej}, \& {Habraken}}]{Delacroix2013}
{Delacroix}, C., {Absil}, O., {Forsberg}, P., {et~al.} 2013, \aap, 553, A98

\bibitem[{{Delacroix} {et~al.}(2012){Delacroix}, {Absil}, {Mawet}, {Hanot},
  {Karlsson}, {Forsberg}, {Pantin}, {Surdej}, \& {Habraken}}]{Delacroix2012}
{Delacroix}, C., {Absil}, O., {Mawet}, D., {et~al.} 2012, \procspie, 8446, 8

\bibitem[{{Dohlen} {et~al.}(2011){Dohlen}, {Wildi}, {Puget}, {Mouillet}, \&
  {Beuzit}}]{Dohlen2011}
{Dohlen}, K., {Wildi}, F.~P., {Puget}, P., {Mouillet}, D., \& {Beuzit}, J.-L.
  2011, in Second International Conference on Adaptive Optics for Extremely
  Large Telescopes., 75

\bibitem[{{Foo} {et~al.}(2005){Foo}, {Palacios}, \& {Swartzlander}}]{Foo2005}
{Foo}, G., {Palacios}, D.~M., \& {Swartzlander}, Jr., G.~A. 2005, Optics
  Letters, 30, 3308

\bibitem[{{Jenkins}(2008)}]{Jenkins2008}
{Jenkins}, C. 2008, \mnras, 384, 515

\bibitem[{{Jovanovic} {et~al.}(2015){Jovanovic}, {Martinache}, {Guyon},
  {Clergeon}, {Singh}, {Kudo}, {Garrel}, {Newman}, {Doughty}, {Lozi}, {Males},
  {Minowa}, {Hayano}, {Takato}, {Morino}, {Kuhn}, {Serabyn}, {Norris},
  {Tuthill}, {Schworer}, {Stewart}, {Close}, {Huby}, {Perrin}, {Lacour},
  {Gauchet}, {Vievard}, {Murakami}, {Oshiyama}, {Baba}, {Matsuo}, {Nishikawa},
  {Tamura}, {Lai}, {Marchis}, {Duchene}, {Kotani}, \&
  {Woillez}}]{Jovanovic2015}
{Jovanovic}, N., {Martinache}, F., {Guyon}, O., {et~al.} 2015, ArXiv e-prints

\bibitem[{{Kerber} {et~al.}(2014){Kerber}, {K{\"a}ufl}, {Baksai}, {Di Lieto},
  {Dobrzycka}, {Duhoux}, {Finger}, {Heikamp}, {Ives}, {Jakob}, {Lundin},
  {Mawet}, {Mehrgan}, {Momany}, {Moreau}, {Pantin}, {Riquelme}, {Sandrock},
  {Siebenmorgen}, {Smette}, {Taylor}, {van den Ancker}, {Valdes}, {Venema}, \&
  {Weilenmann}}]{Kerber2014}
{Kerber}, F., {K{\"a}ufl}, H.-U., {Baksai}, P., {et~al.} 2014, \procspie, 9147,
  0

\bibitem[{Krist {et~al.}(2012)Krist, Belikov, Mawet, Moody, Pueyo, \&
  Shaklan}]{Krist2012}
Krist, J., Belikov, R., Mawet, D., {et~al.} 2012, Technology Milestone 1
  Results Report, JPL Document

\bibitem[{{Mas} {et~al.}(2012){Mas}, {Baudoz}, {Rousset}, \&
  {Galicher}}]{Mas2012}
{Mas}, M., {Baudoz}, P., {Rousset}, G., \& {Galicher}, R. 2012, \aap, 539, A126

\bibitem[{{Mawet} {et~al.}(2013){Mawet}, {Absil}, {Delacroix}, {Girard},
  {Milli}, {O'Neal}, {Baudoz}, {Boccaletti}, {Bourget}, {Christiaens},
  {Forsberg}, {Gonte}, {Habraken}, {Hanot}, {Karlsson}, {Kasper}, {Lizon},
  {Muzic}, {Olivier}, {Pe{\~n}a}, {Slusarenko}, {Tacconi-Garman}, \&
  {Surdej}}]{Mawet2013}
{Mawet}, D., {Absil}, O., {Delacroix}, C., {et~al.} 2013, \aap, 552, L13

\bibitem[{{Mawet} {et~al.}(2011){Mawet}, {Mennesson}, {Serabyn}, {Stapelfeldt},
  \& {Absil}}]{Mawet2011}
{Mawet}, D., {Mennesson}, B., {Serabyn}, E., {Stapelfeldt}, K., \& {Absil}, O.
  2011, \apjl, 738, L12

\bibitem[{{Mawet} {et~al.}(2012){Mawet}, {Pueyo}, {Lawson}, {Mugnier}, {Traub},
  {Boccaletti}, {Trauger}, {Gladysz}, {Serabyn}, {Milli}, {Belikov}, {Kasper},
  {Baudoz}, {Macintosh}, {Marois}, {Oppenheimer}, {Barrett}, {Beuzit},
  {Devaney}, {Girard}, {Guyon}, {Krist}, {Mennesson}, {Mouillet}, {Murakami},
  {Poyneer}, {Savransky}, {V{\'e}rinaud}, \& {Wallace}}]{Mawet2012}
{Mawet}, D., {Pueyo}, L., {Lawson}, P., {et~al.} 2012, \procspie, 8442, 4

\bibitem[{{Mawet} {et~al.}(2005){Mawet}, {Riaud}, {Absil}, \&
  {Surdej}}]{Mawet2005}
{Mawet}, D., {Riaud}, P., {Absil}, O., \& {Surdej}, J. 2005, \apj, 633, 1191

\bibitem[{{Mawet} {et~al.}(2009){Mawet}, {Serabyn}, {Liewer}, {Hanot},
  {McEldowney}, {Shemo}, \& {O'Brien}}]{Mawet2009}
{Mawet}, D., {Serabyn}, E., {Liewer}, K., {et~al.} 2009, Optics Express, 17,
  1902

\bibitem[{{Milli} {et~al.}(2014){Milli}, {Lagrange}, {Mawet}, {Absil},
  {Augereau}, {Mouillet}, {Boccaletti}, {Girard}, \& {Chauvin}}]{Milli2014}
{Milli}, J., {Lagrange}, A.-M., {Mawet}, D., {et~al.} 2014, \aap, 566, A91

\bibitem[{{Noll}(1976)}]{Noll1976}
{Noll}, R.~J. 1976, Journal of the Optical Society of America, 66, 207

\bibitem[{{Reggiani} {et~al.}(2014){Reggiani}, {Quanz}, {Meyer}, {Pueyo},
  {Absil}, {Amara}, {Anglada}, {Avenhaus}, {Girard}, {Carrasco Gonzalez},
  {Graham}, {Mawet}, {Meru}, {Milli}, {Osorio}, {Wolff}, \&
  {Torrelles}}]{Reggiani2014}
{Reggiani}, M., {Quanz}, S.~P., {Meyer}, M.~R., {et~al.} 2014, \apjl, 792, L23

\bibitem[{{Rouan} {et~al.}(2000){Rouan}, {Riaud}, {Boccaletti}, {Cl{\'e}net},
  \& {Labeyrie}}]{Rouan2000}
{Rouan}, D., {Riaud}, P., {Boccaletti}, A., {Cl{\'e}net}, Y., \& {Labeyrie}, A.
  2000, \pasp, 112, 1479

\bibitem[{Sauvage {et~al.}(2012)Sauvage, Mugnier, Paul, \&
  Villecroze}]{Sauvage2012}
Sauvage, J.-F., Mugnier, L., Paul, B., \& Villecroze, R. 2012, Opt. Lett., 37,
  4808

\bibitem[{Serabyn {et~al.}(2010)Serabyn, Mawet, \& Burruss}]{Serabyn2010}
Serabyn, E., Mawet, D., \& Burruss, R. 2010, \nat, 464, 1018

\bibitem[{{Serabyn} {et~al.}(2007){Serabyn}, {Wallace}, {Troy}, {Mennesson},
  {Haguenauer}, {Gappinger}, \& {Burruss}}]{Serabyn2007}
{Serabyn}, E., {Wallace}, K., {Troy}, M., {et~al.} 2007, \apj, 658, 1386

\bibitem[{{Singh} {et~al.}(2014){Singh}, {Martinache}, {Baudoz}, {Guyon},
  {Matsuo}, {Jovanovic}, \& {Clergeon}}]{Singh2014}
{Singh}, G., {Martinache}, F., {Baudoz}, P., {et~al.} 2014, \pasp, 126, 586

\bibitem[{{Wallace} {et~al.}(2010){Wallace}, {Burruss}, {Bartos}, {Trinh},
  {Pueyo}, {Fregoso}, {Angione}, \& {Shelton}}]{Wallace2010}
{Wallace}, J.~K., {Burruss}, R.~S., {Bartos}, R.~D., {et~al.} 2010, \procspie,
  7736, 5

\end{thebibliography}

\appendix 

\section{A Zernike-based analysis}
\label{app:zernike_analysis}

We propose a Fourier-based analysis of beam propagation using Zernike polynomial decomposition of the wavefront. Our computations are based on the standard layout of a coronagraph, illustrated in Fig. \ref{fig:layout}.

\subsection{The Zernike polynomials}
\label{app:zernike-polynomials}

The Zernike polynomials were described by \cite{Noll1976} and are defined for $r \le 1$ ($\vec{r}=(r, \theta)$ are the polar coordinates) as

\begin{equation}
\begin{array}{ll}
Z_{j}(\vec{r})=\sqrt{n+1} R_n^m(r) \sqrt{2}C_m(\theta) &\mathrm{~for~}m \neq  0,  \\
Z_{j}(\vec{r})=\sqrt{n+1}R_n^0(r) & \mathrm{~for~} m=0,
 \end{array}
\end{equation}

\noindent with

\begin{equation}
R_n^m (r) = \sum \limits _ {s=0} ^{(n-m)/2} \dfrac{(-1)^s(n-s)!}{s![(n+m)/2-s]![(n-m)/2-s]!} r^{n-2s}.
\label{eq:Rnm}
\end{equation}

\noindent \corr{Here, }$n$ and $m$ are \corr{non negative} integers satisfying $m \leq n$, with $n - m$ even (in other words, $n$ and $m$ have the same parity). The azimuthal functions, $C_m(\theta) = \cos m \theta$ and $C_m(\theta) = \sin m \theta$, are defined for even and odd $j$ respectively, thus corresponding to real symmetric and antisymmetric modes respectively. The index $j$ is a usual numbering system for the different modes, that will be used in the following developments. There is no equation linking the index $j$ and the $(n,m)$ integer pairs. The correspondences for the first 14 polynomials and their usual aberration designation can be found in Fig.\,\ref{fig:Zernike_conversion}.

The Fourier transform of the Zernike polynomials, noted $\widehat{Z_j}$, can be written as

\begin{equation}
 \renewcommand{\arraystretch}{2}
\begin{array}{ll}
\widehat{Z}_j(\boldsymbol{\alpha})=\sqrt{2 (n+1)} \pi C_m(\psi) i^{-m} (-1)^{\frac{n-m}{2}}\dfrac{2 J_{n+1}(2\pi \alpha)}{2\pi \alpha} &\mathrm{~for~}m \neq 0,  \\
\widehat{Z_{j}}(\boldsymbol{\alpha})=\sqrt{n+1} \pi (-1)^{\frac{n}{2}}\dfrac{2 J_{n+1}(2\pi \alpha)}{2\pi \alpha}  &\mathrm{~for~}m = 0 , \\
\end{array}
\label{eq:zernike_FT}
\end{equation} 

\noindent with $(\alpha, \psi) =  \boldsymbol{\alpha}$ the polar coordinates in the conjugate plane, and $J_n(\alpha)$ the Bessel function of the first kind.

\subsection{The small aberration assumption}
\label{app:small-aberration}

Under the small aberration hypothesis, the wavefront at the entrance pupil (amplitude of $1$, phase $\phi$) can be directly approximated at first order as a combination of Zernike polynomials:

\begin{equation}
\label{eq:Epup}
E_{\rm pup} = \exp(i \phi) \approx 1 + i \phi =  Z_1(\vec{r}) + i  \sum \limits _ {j=2} ^\infty a_j Z_j(\vec{r}),
\end{equation}

\noindent where $\vec{r}=(r, \theta)$ are the polar coordinates in the pupil plane and ${a_j}$ a set of real coefficients describing the aberrations.

The Fourier transform of Eq \ref{eq:Epup} leads to the field distribution in the focal plane and thus consists of a linear combination of Zernike polynomial Fourier transforms, $\widehat{Z_j}$. \corr{We can thus write}

\begin{equation}
E_{\rm foc} = \widehat{ Z_1}(\boldsymbol{\alpha}) + i  \sum \limits _ {j=2} ^\infty a_j \widehat{Z_j}(\boldsymbol{\alpha})
\label{eq:Efoc}
\end{equation}

At the focal plane, the vortex phase mask induces a phase shift depending on the azimuthal angle $\psi$. Indeed, \cite{Mawet2005} have shown that for a perfect vortex phase of topological charge $l_p$, the right- and left-handed circular polarization unit vectors are translated into left- and right-handed circular polarization vectors respectively, and are affected by a phase ramp $ {\rm e}^{ i l_p \psi}$ and $ {\rm e}^{- i l_p \psi}$ respectively. The coronagraphic effect will occur for any value of $l_p$ that is even.

To ease the comparison between the field in entrance pupil ($E_{\rm pup}$) and Lyot plane ($E_{\rm Lyot}$), an inverse Fourier transform, noted $\mathcal{F}^{-1}$, is finally applied, \corr{leading to}

\begin{equation}
\label{eq:Eout}
 \renewcommand{\arraystretch}{2.}
\begin{array}{cccl}
E_{\rm Lyot} & = & \mathcal{F}^{-1} \left[ \widehat{Z}_1(\boldsymbol{\alpha}) {\rm e}^{il_p \psi} \right]& +i \sum \limits_{i=2}^{\infty}a_j \mathcal{F}^{-1} \left[ \widehat{Z}_j(\boldsymbol{\alpha})  {\rm e}^{i l_p \psi} \right] \\
& = &  \zeta _1(\vec{r}) & + i \sum \limits_{j=2}^{\infty}a_j \zeta_j(\vec{r})
\end{array}
\end{equation}

\noindent with $\zeta_j = \mathcal{F}^{-1} \left[ \widehat{Z}_j(\boldsymbol{\alpha}) {\rm e}^{il_p \psi} \right]$ denoting the field distribution in the Lyot plane when the input pupil amplitude is defined by the Zernike polynomial $Z_j$.

The first term $\zeta_1(\vec{r})$ results from the perfect plane component (piston mode) that is completely diffracted outside the geometric pupil in the Lyot plane as long as the charge $l_{\rm p}$ is even \citep{Mawet2005}. In the following section, we derive the general expression of $\zeta_j(\vec{r})$ and show that they can be expressed as Zernike polynomials inside the geometrical pupil.

\subsection{The conversion tables $Z_j \rightarrow \zeta_j$}
\label{app:conversion-table}

\begin{figure*}
\begin{tabular}{lcl}
a) Charge $l_p=2$ &\hspace*{1.2cm}& b) Charge $l_p=4$ \\
\\
\includegraphics[width=.4\linewidth]{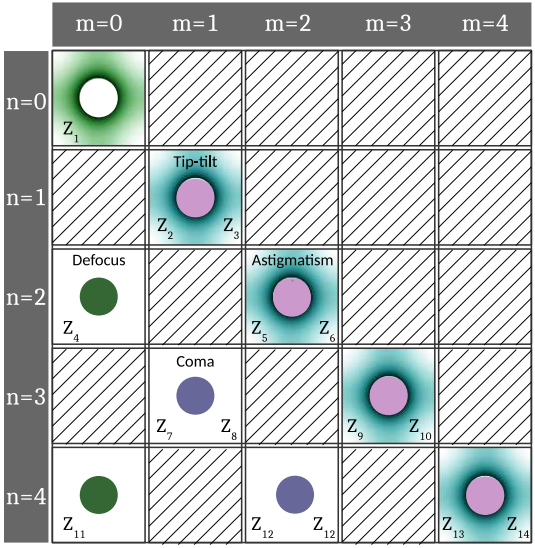} &&
\includegraphics[width=.4\linewidth]{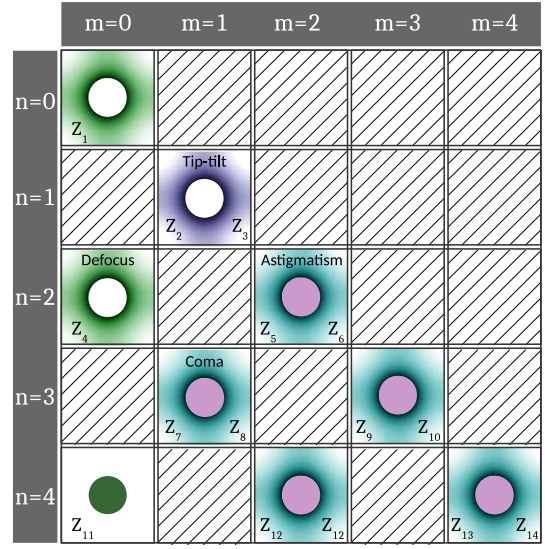}
\end{tabular}
\caption{Qualitative representation of the field distribution at the Lyot plane for a VC of charge $l_p=2$ (left) and $l_p=4$ (right), as a function of the $(n,m)$ integer pair of the input Zernike polynomial $Z_j$. In the column $m=0$, only one term defines the field distribution, and is depicted in green. For all the other cases, two terms contribute to the electric field, represented in cyan and pink. In the case where both terms are superimposed either outside or inside the pupil, these colors are mixed into a purple hue (see for instance the coma for the charge $l_{\rm p}=2$ vortex, or the tip-tilt for the charge $l_{\rm p}=4$ vortex).}
\label{fig:Zernike_conversion}
\end{figure*}

The expression of $\zeta_j$ can be expanded using Eq.\,\ref{eq:zernike_FT}, \corr{thus becoming}

\begin{equation}
\label{eq:Zout}
\zeta_j(\vec{r}) = A_n^m  \int_0^{\infty} \dfrac{2 J_{n+1}(2\pi \alpha)}{2\pi \alpha} \int_0^{2\pi}C_m {\rm e}^{il_p \psi} {\rm e}^{i2 \pi r \alpha \cos(\theta-\psi)} \alpha {\rm d} \alpha {\rm d} \psi,
\end{equation}

\noindent with 

\begin{equation}
\begin{array}{ll}
A_n^m= \sqrt{2 (n+1)}\pi(-1)^{\frac{n-m}{2}}i^{-m} & \text{ for } m \neq 0, \\
A_n^0= \sqrt{n+1}\pi(-1)^{\frac{n}{2}} & \text{ for } m = 0,
\end{array}
\end{equation}

\noindent and

\begin{equation}
C_m  = 
\begin{cases}
\cos(m\psi) = \frac{1}{2} ({\rm e}^{i m \psi} + {\rm e}^{-i m \psi})  &\text{ for }m \neq 0 \text{ and even }j, \\
\sin(m\psi)  = \frac{-i}{2} ({\rm e}^{i m \psi} - {\rm e}^{-i m \psi})  &\text{ for }m \neq 0 \text{ and odd }j, \\
1 &\text{ for }m=0.
\end{cases}
\end{equation}

\noindent $\zeta_j(\vec{r})$ will thus be written as one term ($m=0$) or as the sum of two terms ($m \neq 0$). In any case, all these terms have the same form and can be simplified using the integral form of the Bessel function \corr{that expresses as}

\begin{equation}
J_k(z)=\frac{1}{2 \pi i^{k}}\int_{0}^{2\pi} {\rm e}^{ik\varphi} {\rm e}^{iz\cos(\varphi)} {\rm d}\varphi,
\end{equation}

\noindent which can also be written in the more convenient manner

\begin{equation}
\int_{0}^{2\pi} {\rm e}^{i k \psi} {\rm e}^{iz\cos( \theta-\psi)} {\rm d} \psi =2 \pi i^{k} {\rm e}^{i k \theta} J_k(z),
\end{equation}

\noindent where we identify \corr{$\correq{z=2 \pi r \alpha}$, and} $k=l_p$ when $m = 0$ or $k=l_p \pm m$ when $m \neq 0$.

Replacing the integral over the variable $\psi$, the general expression of $\zeta_j(\vec{r})$ becomes

\begin{equation}
\zeta_j(\vec{r}) = A_n^m \times
 \begin{cases}
~~ (I_n^{l_p+m} + I_n^{l_p-m}) &\text{ for }m \neq 0 \text{ and even }j, \\
-i (I_n^{l_p+m} - I_n^{l_p-m}) &\text{ for }m \neq 0 \text{ and odd }j, \\
 ~~ 2 I_n^{l_p}&\text{ for }m = 0,  \\
\end{cases}
\label{eq:zetai_Ink}
\end{equation}

\noindent with

\begin{equation}
I_n^k =  i^k {\rm e}^{i k \theta} \int_0^{\infty}  J_{n+1}(2\pi \alpha)  J_{k}(2 \pi \alpha r)   {\rm d} \alpha.
\label{eq:disc_int}
\end{equation}

According to the Eq. 9 of \cite{Noll1976}, the integral can be linked to the $R_n^m(r)$ function (recalled in Eq.\,\ref{eq:Rnm}), that we rewrite here in a more general manner (not restricted to the conditions $k \leq n$ and $k$ and $n$ with the same parity):

\begin{equation}
I_n^k =  i^k {\rm e}^{i k \theta}
  \begin{cases}
    \text{ for }|k| \leq n \begin{cases}	
    					 \frac{1}{2 \pi}(-1)^\frac{k-n}{2}R_n^{|k|}(r) ~~~~~~~~~~~ &\text{ for }  0 \leq r \leq 1 	\\
					0 & \text{ for }r > 1
			\end{cases} \\
     \text{ for }|k|> n 	\begin{cases}	
    					 0  & \text{ for }0 \leq r \leq 1 	\\
					-\frac{1}{2 \pi r}(-1)^\frac{n-k}{2}R_{|k|-1}^{n+1}(1/r) &\text{ for }r > 1			
		\end{cases} \\
  \end{cases}
\end{equation}

\noindent One can note that this integral is discontinuous. The general form of $\zeta_j(\vec{r})$ can thus be considered as two components that are non-zero exclusively inside or outside the geometrical pupil, depending on the comparison of $\vert l_p \pm m \vert $ with $n$. \corr{We can thus write }

\begin{equation}
\zeta_j = \zeta_j^{\rm in} + \zeta_j^{\rm out},
\end{equation} 

\noindent where $\zeta_j^{\rm in}$ and $\zeta_j^{\rm out}$ represent the contributions of the field inside and outside the pupil respectively. This field distribution is visually illustrated in Fig.\,\ref{fig:Zernike_conversion} for a VC of charge $l_p=2$ and $l_p=4$. Note that $\zeta_j^{\rm in}$ or $\zeta_j^{\rm out}$ can be zero. Indeed, if $m=0$, there is only one term (Eq.\,\ref{eq:zetai_Ink}), which is defined either for $r > 1$ or $0 \leq r \leq 1$. In particular for the piston term (plane wavefront), there is no component inside the geometrical pupil, confirming the theoretical perfect extinction of the VC. Another interesting example is the case of the defocus and tip-tilt modes: in the case of the VC of charge $l_p=4$, both of them fall outside the geometrical pupil, while there is a non zero contribution inside the pupil for the VC of charge $l_p=2$. This result confirms that, for circular unobstructed pupils, charge 4 vortices are less sensitive to tip-tilt and defocus aberrations than charge 2 vortices, since at first order, these modes are completely rejected outside the pupil.

The final expression of $\zeta_j(\vec{r})$ thus depends on the value of $m$ and the parity of $j$. As a summary, we can write:

\begin{itemize}
\item when $m=0$, it implies that $n$ is even and hence:
\begin{equation}
\renewcommand{\arraystretch}{2.}
\zeta_j(\vec{r})=
\begin{cases}
\sqrt{n+1} {\rm e}^{i l_p \theta} R_n^{|l_p|}(r)  & \text{ for } |l_p| \leq n, 0 \leq r \leq 1, \\ 
-\sqrt{n+1} \frac{1}{r}{\rm e}^{i l_p \theta} R_{|l_p|-1}^{n+1}(1/r) & \text{ for } |l_p| > n, r > 1.
\end{cases}
\end{equation}

\item when $m \neq 0$, we distinguish the cases even and odd $j$:

\begin{equation}
\zeta_j(\vec{r})= 
\begin{cases}
	T^{l_p+m}_n + T^{l_p-m}_n &\text{ for even } j, \\
	-i T^{l_p+m}_n + i T^{l_p-m}_n &\text{ for odd } j. \\
\end{cases}
\end{equation}

\noindent with 

\begin{equation}
T_{n}^{k} = 
\begin{cases}
\sqrt{\frac{n+1}{2}}  {\rm e}^{ ik \theta} R_n^{|k|}(r)  & \text{ for } |k| \leq n, 0 \leq r \leq 1, \\
 -\sqrt{\frac{n+1}{2}} \dfrac{{\rm e}^{ i k \theta}}{r} R_{|k|-1}^{n+1} \left(\frac{1}{r} \right)  & \text{ for } |k| > n, r > 1 .
\end{cases}
\label{eq:Tnk}
\end{equation}

\end{itemize}

It is interesting to note that, if they exist, the terms that are defined inside the geometrical pupil ($r \leq 1$) can be expressed as a complex combination of Zernike polynomials. \corr{We can indeed write}

\begin{equation}
{\rm e}^{ ik \theta} R_n^{|k|}(r) = \correq{ \frac{1}{\sqrt{n+1}} } 
\begin{cases}
\left( Z_{\text{even }j}^{n,k} \pm i  Z_{\text{odd }j}^{n,k} \right)\correq{ / \sqrt{2} } &\text{ if } k = l_p\pm m \neq 0, \\
Z_{j}^{n,0} & \text{ if } k = l_p \pm m = 0,
\end{cases}
\end{equation}

\noindent with the $\pm$ sign corresponding to the sign of $k$. As a consequence, a conversion table can be established, that gives the coefficients of the Zernike polynomials defining the field after the Lyot stop (i.e. $\zeta_j^{\rm in}$, since the  $\zeta_j^{\rm out}$ is blocked by the aperture stop) for a given input Zernike polynomial, $Z_j$, passing through the VC. These tables are given for the charge $l_{\rm p}=2$ and $l_{p}=4$ vortices (Table \ref{tab:charge2} and \ref{tab:charge4}). 

\begin{table}
\caption{Conversion table for the first eight Zernike polynomials for a charge $l_p=2$ vortex phase mask. Only the contribution inside the geometrical pupil is considered (electric field after the Lyot stop). As a reminder, $\zeta_j^{\rm in}$ corresponds to the contribution of the input Zernike polynomial $Z_j$ inside the geometrical pupil, such that the table should be read line by line (for instance, if the entrance pupil contains the tip-tilt mode $Z_2$, this will translate in the Lyot plane as $\zeta_2^{\rm in} = Z_2/2+i Z_3/2$).}
\label{tab:charge2}
\renewcommand{\arraystretch}{1.5}
\begin{tabular}{cccccccccc}
\hline \hline
             	& 	$Z_2$ 	&	$Z_3$	&	$Z_4$	& $Z_5$ 	&	$Z_6$	&	$Z_7$	&	$Z_8$	&	$Z_9$	&	$Z_{10}$ \\
\hline
$\zeta_1^{\rm in}=$ 	&	0			& 0			&	0			&	0			&	0			&	0			&	0			&	0			&	0				\\

$\zeta_2^{\rm in}=$ 	&	$\frac{1}{2}$ &$\frac{i}{2}$&	0			&	0			&	0			&	0			&	0			&	0			&	0				\\

$\zeta_3^{\rm in}=$ 	&	$\frac{i}{2}$& $-\frac{1}{2}$	&	0			&	0			&	0			&	0			&	0			&	0			&	0				\\

$\zeta_4^{\rm in}=$ 	&	0			& 0			&	0			&	$\frac{i}{\sqrt{2}}$	&	$\frac{1}{\sqrt{2}}$&	0			&	0			&	0			&	0				\\

$\zeta_5^{\rm in}=$ 	&	0			& 0			&	$\frac{i}{\sqrt{2}}$	&	0			&	0			&	0			&	0			&	0			&	0				\\

$\zeta_6^{\rm in}=$ 	&	0			& 0			&	$\frac{1}{\sqrt{2}}$ &	0			&	0			&	0			&	0			&	0			&	0				\\

$\zeta_7^{\rm in}=$ 	&	0			& 0			&	0			&	0			&	0			&	$-\frac{1}{2}$&$\frac{i}{2}$&$\frac{1}{2}$&$-\frac{i}{2}$\\

$\zeta_8^{\rm in}=$ 	&	0			& 0			&	0			&	0			&	0			&	$\frac{i}{2}$&$\frac{1}{2}$&$\frac{i}{2}$&$\frac{1}{2}$\\
\hline
\end{tabular}
\end{table}

\begin{table}
\caption{Same as table \ref{tab:charge2}, but for a charge $l_p=4$ vortex phase mask.}
\label{tab:charge4}
\renewcommand{\arraystretch}{1.5}
\begin{tabular}{cccccccccc}
\hline \hline
             	& 	$Z_2$ 	&	$Z_3$	&	$Z_4$	& $Z_5$ 	&	$Z_6$	&	$Z_7$	&	$Z_8$	&	$Z_9$	&	$Z_{10}$ \\
\hline
$\zeta_1^{\rm in}=$ 	&	0			& 0			&	0			&	0			&	0			&	0			&	0			&	0			&	0				\\

$\zeta_2^{\rm in}=$ 	&	0			 &	0			&	0			&	0			&	0			&	0			&	0			&	0			&	0				\\

$\zeta_3^{\rm in}=$ 	&	0			&	0			&	0			&	0			&	0			&	0			&	0			&	0			&	0				\\

$\zeta_4^{\rm in}=$ 	&	0			& 0			&	0			&	0			&	0			&	0			&	0			&	0			&	0				\\

$\zeta_5^{\rm in}=$ 	&	0			& 0			&	0			&	-$\frac{1}{2}$ & $\frac{i}{2}$ &	0			&	0			&	0			&	0				\\

$\zeta_6^{\rm in}=$ 	&	0			& 0			&	0			 & $\frac{i}{2}$ & $\frac{1}{2}$	&	0			&	0			&	0			&	0				\\
$\zeta_7^{\rm in}=$ 	&	0			& 0			&	0			&	0			&	0			&	0			& 0			& $-\frac{1}{2}$	& $\frac{i}{2}$	\\

$\zeta_8^{\rm in}=$ 	&	0			& 0			&	0			&	0			&	0			&0			&	0			&$\frac{i}{2}$&$\frac{1}{2}$ \\
\hline
\end{tabular}
\end{table}

\section{Off-axis transmission}

\subsection{Analytical function}

The Zernike analysis that has been carried for a tilted wavefront (Eq.\,\ref{eq:Elyot_ch2}) allows the estimation of the transmission efficiency for an off-axis source close to the center. Because of the central symmetry, only one axis is needed to describe the transmission as a function of the distance from the axis, noted $T$ in rad rms. At the first order, the total transmission is estimated from Eq.\,\ref{eq:Elyot_ch2} by

\begin{equation}
\eta_{l_p=2}(T) = \dfrac{\int_{\rm pup} \vert E_{\rm Lyot} \vert^2}{\int_{\rm pup} \vert E_{\rm pup} \vert^2} = \dfrac{T^2}{2}.
\end{equation}

\noindent For convenience, the tip-tilt rms in radian, $T$, can be converted into an amplitude $S$ in unit of $\lambda/D$ by means of the relation $T_{[\mathrm{rad}]}=S_{[\lambda/D]} \times \pi/2$, leading to

\begin{equation}
\begin{array}{cc}
\eta_{l_p=2} \left( S \right) &=\dfrac{\pi^2}{8}  S_{[\lambda/D]}^2. \\
\end{array}
\label{eq:eta}
\end{equation}

\begin{figure}
\centering
\includegraphics[width=.9\linewidth]{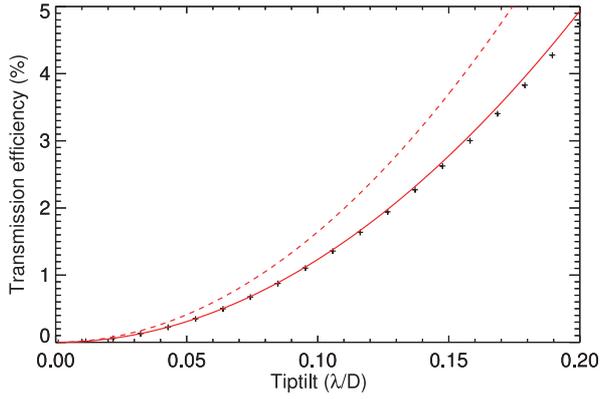}
\caption{Transmission efficiency as a function of the angular distance from the center. The crosses result from simulations while the dashed line shows the theoretical function as stated by \cite{Jenkins2008}, and the solid line shows the model derived in this work. The two are in agreement for the square dependency, but differ in the multiplicative factor, equalling $\pi^2/6$ and $\pi^2/8$ respectively.}
\label{fig:transmission_vs_tiptilt}
\end{figure}

This result is slightly different from the formula given by \cite{Jenkins2008}, who derived it empirically. His result is also based on a square law but the multiplicative factor is different ($\pi^2/6$ instead of $\pi^2/8$). Simulations have been performed in order to compare the two models. The parameters of the simulations are: a grid size of 1024 points in width, entrance pupil covering 102 pixels and a Lyot stop of the same size as the entrance pupil. Particular care has to be given to numerical errors: the main part of them can be avoided by computing the entrance pupil profile that leads to perfect attenuation of an on-axis source. This is performed by simulating the propagation of a perfect circular wavefront up to the Lyot plane, cancelling out the residuals inside the geometrical pupil (relying on the argument that this is true analytically), and finally propagating the result backwards, down to the entrance pupil \citep{Krist2012}. The complex profile of the entrance pupil obtained in this way is used as the perfect wavefront. The results of the tip-tilt simulations are shown in Fig. \ref{fig:transmission_vs_tiptilt} and confirm that for small tip-tilt values, the transmission efficiency follows the function given in Eq. \ref{eq:eta}.

\subsection{Experimental results}
\label{app:off-axis_exp}

\begin{figure}
\centering
\includegraphics[width=.9\linewidth]{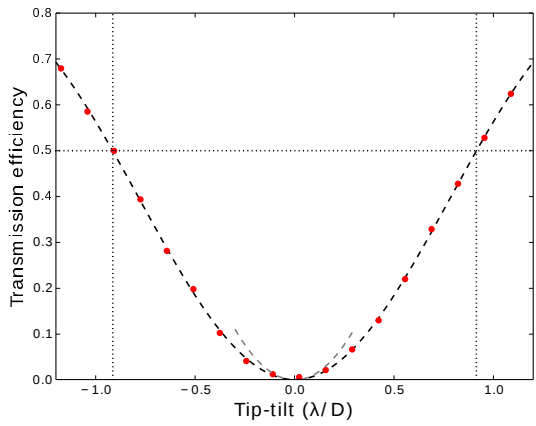}
\caption{Measured transmission efficiency for an off-axis source through a VC. The red circles correspond to experimental results, while the dark dashed line is the best fit model (polynomial function of even orders only, up to the 6th order). The light gray dashed line corresponds to the theoretical model as stated by Eq. \ref{eq:eta}, assuming very small tip-tilt (it is thus drawn only for absolute tip-tilt $<0.3\,\lambda/D$). The inner working angle is graphically represented by the dotted lines, that highlight the transmission limit of 50\%, reached for tip-tilt of $0.9\,\lambda/D$.}
\label{fig:transmission_results}
\end{figure}

The experimental data described in Sect. \ref{sec:experimental_validation} have been processed in order to estimate the transmission efficiency as a function of tip-tilt. The flux has been integrated for each position of tip-tilt in a square of width $10\,\lambda/D$ centred on the PSF, and divided by the value obtained for the AGPM translated by $7\,\lambda/D$, a distance at which the beam is barely affected by the vortex phase mask. The transmission curve is shown in Fig. \ref{fig:transmission_results}. A polynomial function (composed only of even orders up to the $6^{\rm th}$, because of the obvious and expected symmetry) has been fitted to the data points. The best fit model leads to a position of the minimal transmission around $-0.02\,\lambda/D$, meaning that the position that was thought to be the optimal was actually off by $3.5\,\muup$m in the focal plane. The inner-working angle, defined as the distance where the off-axis transmission reaches 50\%, is estimated to be 0.9$\,\lambda/D$ (with $D$ the diameter of the entrance pupil). The results were also compared to the theoretical model as derived in Eq. \ref{eq:eta}, but the sampling was obviously not sufficient at very small tip-tilt to perform a useful comparison.

\end{document}